\documentclass[longauth]{aa} 
\usepackage[varg]{txfonts}
\usepackage{graphicx,epsf}
\usepackage{amsmath,graphicx,adjustbox,setspace, sidecap, float}
\usepackage[toc,page]{appendix}
\usepackage{gensymb}
\usepackage{hyperref}
\usepackage{fancyhdr}
\usepackage{natbib}
\usepackage{caption,subcaption}
\usepackage{xcolor}
\usepackage{physics}
\usepackage{ulem}
\usepackage{multirow}
\usepackage{cancel}
\usepackage{xspace}

\usepackage[para]{threeparttable}
\usepackage{orcidlink}

\newcommand{\sect}[1]{Sect.~\ref{sec:#1}}
\newcommand{\Sect}[1]{Section~\ref{sec:#1}}

\newcommand{\fg}[1]{Fig.~\ref{fig:#1}}
\newcommand{\Fg}[1]{Figure~\ref{fig:#1}}

\begin{document}

\title{Dust mineralogy and variability of the inner PDS 70 disk}
\subtitle{Insights from JWST/MIRI MRS and Spitzer IRS observations}

   \author{Hyerin Jang \orcidlink{0000-0002-6592-690X} \inst{1}
   \and 
   Rens Waters \orcidlink{0000-0002-5462-9387} \inst{1,2}
   \and
   Till Kaeufer \orcidlink{0000-0002-5430-1170} \inst{2,3,4,5}
   \and
   Akemi Tamanai \inst{6}
   \and
   Giulia Perotti \orcidlink{0000-0002-8545-6175} \inst{7}
   \and
   Valentin Christiaens \orcidlink{0000-0002-0101-8814} \inst{8,9}
   \and
   Inga Kamp \orcidlink{0000-0001-7455-5349} \inst{4}
   \and
   Thomas Henning \orcidlink{0000-0002-1493-300X} \inst{7}
   \and 
   Michiel Min \inst{2}
   \and
   Aditya M. Arabhavi \orcidlink{0000-0001-8407-4020} \inst{4}
   \and 
   David Barrado \orcidlink{0000-0002-5971-9242} \inst{10}
   \and
   Ewine F. van Dishoeck \orcidlink{0000-0001-7591-1907} \inst{11,12}
   \and
   Danny Gasman \orcidlink{0000-0002-1257-7742} \inst{8}
   \and
   Sierra L. Grant \orcidlink{0000-0002-4022-4899} \inst{12}
   \and
   Manuel G\"udel \orcidlink{0000-0001-9818-0588} \inst{13,14}
   \and
   Pierre-Olivier Lagage \inst{15}
   \and
   Fred Lahuis \inst{16}
   \and
   Kamber Schwarz \orcidlink{0000-0002-6429-9457} \inst{7}
   \and
   Beno\^{i}t Tabone \inst{17}
   \and 
   Milou Temmink \orcidlink{0000-0002-7935-7445} \inst{11} 
   }
   
   \institute{Department of Astrophysics/IMAPP, Radboud University, PO Box 9010, 6500 GL Nijmegen, The Netherlands 
   \newline\email{hyerin.jang@astro.ru.nl}
   \and 
   SRON Netherlands Institute for Space Research, Niels Bohrweg 4, NL-2333 CA Leiden, the Netherlands 
   \and  
   Space Research Institute, Austrian Academy of Sciences, Schmiedlstrasse 6, 8042 Graz, Austria 
   \and
   Kapteyn Astronomical Institute, Rijksuniversiteit Groningen, Postbus 800, 9700AV Groningen, The Netherlands 
   \and
   Institute for Theoretical Physics and Computational Physics, Graz University of Technology, Petersgasse 16, 8010 Graz, Austria 
   \and
   RIKEN Cluster for Pioneering Research, 2-1 Hirosawa, Wako-shi, Saitama 351-0198, Japan 
   \and 
   Max-Planck-Institut f\"{u}r Astronomie (MPIA), K\"{o}nigstuhl 17, 69117 Heidelberg, Germany 
   \and 
   Institute of Astronomy, KU Leuven, Celestijnenlaan 200D, 3001 Leuven, Belgium 
   \and 
   STAR Institute, Universit\'e de Li\`ege, All\'ee du Six Ao\^ut 19c, 4000 Li\`ege, Belgium 
   \and
   Centro de Astrobiolog\'ia (CAB), CSIC-INTA, ESAC Campus, Camino Bajo del Castillo s/n, 28692 Villanueva de la Ca\~nada, Madrid, Spain 
   \and 
   Leiden Observatory, Leiden University, 2300 RA Leiden, the Netherlands 
   \and
   Max-Planck Institut f\"{u}r Extraterrestrische Physik (MPE), Giessenbachstr. 1, 85748, Garching, Germany 
   \and
   Dept. of Astrophysics, University of Vienna, T\"urkenschanzstr. 17, A-1180 Vienna, Austria 
   \and
   ETH Z\"urich, Institute for Particle Physics and Astrophysics, Wolfgang-Pauli-Str. 27, 8093 Z\"urich, Switzerland 
   \and
   Universit\'e Paris-Saclay, Universit\'e Paris Cit\'e, CEA, CNRS, AIM, F-91191 Gif-sur-Yvette, France 
   \and
   SRON Netherlands Institute for Space Research, PO Box 800, 9700 AV, Groningen, The Netherlands 
   \and 
   Universit\'e Paris-Saclay, CNRS, Institut d’Astrophysique Spatiale, 91405, Orsay, France 
   }
        
\abstract
{
\textit{Context.} The inner disk of the young star PDS~70 may be a site of rocky planet formation, with two giant planets detected further out. Recently, JWST/MIRI MRS observations have revealed the presence of warm water vapour in the inner disk. Solids in the inner disk may inform us about the origin of this inner disk water and nature of the dust in the rocky planet-forming regions of the disk. \\
\textit{Aims.} We aim to constrain the chemical composition, lattice structure, and grain sizes of small silicate grains in the inner disk of PDS~70, observed both in JWST/MIRI MRS and Spitzer IRS.\\
\textit{Methods.} We use a dust fitting model, called DuCK, based on a two-layer disk model considering three different sets of dust opacities. We use Gaussian Random Field and Distribution of Hollow Spheres models to obtain two sets of dust opacities using the optical constants of cosmic dust analogues derived from laboratory-based measurements. These sets take into account the grain sizes as well as their shapes. The third set of opacities is obtained from the experimentally measured transmission spectra from aerosol spectroscopy. We use stoichiometric amorphous silicates, forsterite, and enstatite in our analysis. We also study the iron content of crystalline olivine using the resonance at 23-24~$\mu$m and test the presence of fayalite. Both iron-rich and magnesium-rich amorphous silicate dust species are also employed to fit the observed spectra. \\
\textit{Results.} The Gaussian Random Field opacity set agrees well with the observed spectrum, better than the other two opacity sets. In both MIRI and Spitzer spectra, amorphous silicates are the dominant dust species. Crystalline silicates are dominated by iron-poor olivine. The 23-24~$\mu$m olivine band peaks at 23.44 $\mu$m for the MIRI spectrum and 23.47 $\mu$m for the Spitzer spectrum, representing around or less than 10 \% of iron content in the crystalline silicate. In all of models, we do not find strong evidence for enstatite. Moreover, the silicate band in the MIRI spectrum indicates larger grain sizes (a few microns up to 5 $\mu$m) than the Spitzer spectrum (0.1 to 1~$\mu$m), indicating a time-variable small grain reservoir.  \\
\textit{Conclusions.} The inner PDS~70 disk is dominated by a variable reservoir of warm (T $\sim$350-500~K) amorphous silicates, with $\sim$15 \% of forsterite in mass fraction. The 10$\mu$m and 18$\mu$m amorphous silicate bands are very prominent, indicating that most emission originates from optically thin dust. We suggest that the small grains detected in the PDS~70 inner disk are likely transported inward from the outer disk as a result of filtration by the pressure bump associated with the gap and fragmentation into smaller sizes at the ice line. Collisions among larger parent bodies may also contribute to the small grain reservoir in the inner disk, but these parent bodies must be enstatite-poor. In addition, the variation between MIRI and Spitzer data can be explained by a combination of grain growth over 15 years and a dynamical inner disk where opacity changes occur resulting from the highly variable hot (T$\sim$1000~K) innermost dust reservoir.
}

\maketitle
\section{Introduction}
\label{sec:intro}
\paragraph{}
All young solar-type stars are surrounded by a gas-rich and dusty disk in which planets form, and the dust constitutes the fundamental building material for rocky planets in the inner disk \citep{Raymond_Morbidelli2022}. Thus, a better understanding of dust properties in the inner disk is crucial to reveal the evolutionary process of rocky planets which may lead to the development of habitable worlds. At wavelengths shorter than near infrared, observing the dust within a protoplanetary disk is challenging due to the intense flux from the central star. However, at mid-infrared wavelengths, warm dust grains in the inner disk are the main contributors of emission spectra. 

PDS~70 is characterized as a K7 spectral type star, and its disk contains a highly dust-depleted gap between the inner and outer disk with a $\sim$54~au separation \citep{Hashimoto_etal2015,Hashimoto_etal2012,Long_etal2018,Keppler_etal2018}. Analysis of continuum observations from the Atacama Large Millimeter/submillimeter Array (ALMA) constrains the dust mass of the inner disk ($<18$ au) to 0.08–0.36 M$_{\oplus}$ \citep{Benisty_etal2021}. Within the gap, two giant protoplanets,  PDS~70b and PDS~70c, have been detected \citep{Keppler_etal2018,Muller_etal2018,Wagner_etal2018,Haffert_etal2019,Isella_etal2019}, and there is strong evidence that these planets are still accreting \citep{Aoyama_Ikoma2019,Haffert_etal2019,Thanathibodee_etla2019,Hashimoto_etal2020,Zhou_etal2021,CampbellWhite_etal2023}. Recently, James Webb Space Telescope (JWST; \citealt{Rigby_etal2023}) Near Infrared Camera Instrument and Near Infrared Imager and Slitless Spectrograph observations of the PDS~70 disk have detected the two protoplanets, with a circumplanetary disk around PDS~70c, and a potential third protoplanet at around 13.5 au distance from the star \citep{Mesa_etal2019, Christiaens_etal2024,Blakely_etal2024}. The PDS~70 system has also been observed with the JWST Mid-Infrared Instrument (MIRI; \citealt{Rieke_etal2015, Wright_etal2015,Wright_etal2023}) by the MIRI mid-INfrared Disk Survey (MINDS; \citealt{Henning_etal2024}). \cite{Perotti_etal2023} report the detection of warm water vapour ($\sim$600 K) in the inner disk. By using a two-layer disk model, the authors define the dust continuum and identify crystalline and amorphous silicate emission bands.

Silicate stands as the predominant material of dust in both the interstellar medium and protoplanetary disks \citep{Dorschner_etal1995, Colangeli_etal2003,Henning2010}. Silicates in the interstellar medium are mainly amorphous \citep{Kemper_etal2004,Chiar_Tielens2006} whereas the presence of crystalline silicates is confirmed in protoplanetary disks \citep{Bouwman_etal2001, vanBoekel_etal2004, Forrest_etal2004, vanBoekel_etal2005,KesslerSilacci_etal2005,Apai_etal2005, KesslerSilacci_etal2006, Olofsson_etal2009, Sargent_etal2009, Watson_etal2009, SiciliaAguilar_etal2009, Juhasz_etal2010,Olofsson_etal2010, Oliveira_etal2011, Sturm_etal2013}. The formation of crystalline silicates involves either gas phase condensation or thermal annealing of amorphous silicates \citep{Bradley_etal1983}. Therefore, crystalline silicates in disks can trace the thermal history of dust together with amorphous silicates. The dominant dust components are magnesium-iron (Mg-Fe) silicates for both amorphous and crystalline forms. Particularly, crystalline olivine (Mg$_{2x}$Fe$_{2-2x}$SiO$_4$ ($0\leq x \leq1$)) and crystalline pyroxene (Mg$_x$Fe$_{1-x}$SiO$_3$ ($0\leq x \leq1$)) have been detected in various astronomical environments. Likewise, amorphous Mg-Fe silicates having the same chemical formula as mentioned above are one of the most abundant components of dust grains. Identifying these dust species in protoplanetary disks is possible through the shapes of their opacities at infrared wavelengths, that reflect the chemical composition, lattice structure, grain size, and shape of dust grains.

Infrared spectroscopy can be used to study the spatial distribution of crystalline silicates, because of the wide wavelength range accessible and the availability of resonances that probe both warm and cold dust \citep{Juhasz_etal2010,Olofsson_etal2009,Olofsson_etal2010,Sturm_etal2013,Espaillat_etal2011}. For instance, \cite{Juhasz_etal2010} measured the mass fraction of forsterite (crystalline Mg$_2$SiO$_4$) and enstatite (crystalline MgSiO$_3$) at two different wavelength ranges for Herbig Ae/Be stars observed with the Spitzer Space Telescope (Spitzer) InfraRed Spectragraph (IRS): in the wavelength range of 5-17  and 17-35 $\mu$m, representing inner and outer disk, respectively. From the comparison, they found that forsterite is more abundant than enstatite in the outer disk, and vice versa for the inner disk. \cite{Olofsson_etal2009} and \cite{Olofsson_etal2010} investigated crystalline silicates in Spitzer IRS spectra of protoplanetary disks around a large sample of T-Tauri stars. In both warm (emitting at $\lambda\sim$10 $\mu$m) and cold (emitting at $\lambda$>20 $\mu$m) regions in the disks, crystalline silicate bands of forsterite, enstatite, and diopside(CaMgSi2O6) were detected. The crystallinity of the inner and outer disk was found to be roughly similar at average values of 16 and 19 \%, respectively, however with some disks showing a higher outer disk crystallinity compared to the inner disk. Moreover, \cite{Sturm_etal2013} analyzed the 69 $\mu$m forsterite band in Herbig Ae/Be and T-Tauri disks using Herschel. Sources that show the 69 $\mu$m band are well fitted with Mg-rich forsterite, with an upper limit to the iron content of 2-4 \%; the forsterite was found to be at 100-200 K,  i.e. far out in the disk beyond $\sim$10~au. 

The distinctive shapes of band features in opacities of silicate dust species enable us to identify specific silicate species within the disk. Dust opacities can be calculated for a range of grain sizes by adopting a grain shape model. One such theoretical model is the Distribution of Hollow Sphere (DHS; \citealt{Min_etal2005}). DHS provides a distribution for the fraction of empty space within a hollow sphere and measures their opacity. Another theoretical model, the Gaussian Random Field (GRF; \citealt{Min_etal2007}), generates individual grain surface roughness based on a Gaussian random field and calculates the average opacity over a range of the individual shapes of the porous GRF particle. \cite{Min_etal2007} employed the Discrete Dipole Approximation (DDA) to compute the opacity of the porous GRF particles and selected particles containing 2000-3500 dipoles. For more details on this method, we refer to \cite{Min_etal2006, Min_etal2007}. 

Dust opacities can also be directly measured from aerosol particles. This method has the advantage that it avoids theoretical assumptions about grain shape and size \citep{Tamanai_etal2006,Tamanai_etal2009}. In the aerosol experiments, micron-sized aerosol of dust species are measured with a Fourier transformation infrared spectrometer under a condition close to vacuum. This setup ensures that the grain shapes and their environment could closely mimic those of dust in the universe. Given that the shapes of opacity curves are influenced by measurement methods, it becomes crucial to use appropriate opacity curves for accurate dust fitting. 

In this study, we use these dust opacities to analyze the inner-disk dust of the PDS~70 disk observed with  JWST/MIRI Medium Resolution Spectrometer (MRS; \citealt{Argyriou_etal2023}) and Spitzer IRS. We focus on the quantitative analysis of silicate dust bands within the PDS~70 disk. Specifically, we aim to reveal dust species, grain sizes, and their mass abundances. In \sect{obser_reduc}  we introduce the JWST/MIRI MRS observations and the data reduction, and we present the Spitzer IRS observations. In \sect{dustfittingtool}, we describe the dust fitting model, DuCK, that we use to analyze dust bands in the spectrum. The results from fitting with different sets of opacities are shown in \sect{results}, and we also investigated fraction of iron in silicate dust in \sect{ironSilicate}. \Sect{fittingSpitzer} shows fitting results of the Spitzer spectrum for comparison to the MIRI spectrum. We discuss our fitting results, variability between the MIRI and Spitzer spectra, and origin of the inner-disk small grains in \sect{discussion}. Conclusions are summarized in \sect{conclusion}.

\section{Spitzer IRS and JWST/MIRI MRS observations and data reduction}
\label{sec:obser_reduc}
\paragraph{}
The PDS~70 disk has been observed with both Spitzer IRS and JWST/MIRI MRS with these observations being 15 years apart. With Spitzer IRS, the observation was conducted in low-resolution ($R\sim 60-100$) in 2007 as a part of program 40679 (PI: G. Rieke), and we use its reduced data, introduced in \cite{Perotti_etal2023}. These spectra are shown together with the JWST/MIRI MRS observation in \fg{MIRI_Spitzer}. The Spitzer spectrum shows a spike at 14 $\mu$m due to imprecise stitching of the data, but its peak and shape do not match with any of dust bands we have. Thus, the spike does not affect to the dust fitting process. 
    \subsection{JWST/MIRI MRS}
    \paragraph{}
    PDS~70 was observed on August 1, 2022 with the MRS of the MIRI instrument, having spectral resolving power $R \sim 1,600-3,400$. This observation was part of the Guaranteed Time Observation (GTO) program 1282 (PI: Th.Henning), called MINDS \citep{Henning_etal2024}. The observational strategy and data reduction are already explained in \cite{Perotti_etal2023}, so we summarize the overall process and slight differences in the process.
        
    In short, the observed data were reduced using the MINDS pipeline (v1.0.0; \citealt{Christiaens_etal2024ascl}), a hybrid pipeline leveraging on the one hand the JWST Science Calibration pipeline \citep[version 1.12.5;][]{Bushouse_etal2023} along with the Calibration Reference Data System (CRDS) context jwst\_1118.pmap, and on the other hand the Vortex Image Processing \citep[VIP;][]{Gomez_etal2017, Christiaens_etal2023} package for tasks such as bad pixel correction, background subtraction, and source centroid finding. 
    
    In Detector1 and Spec2 stages of the pipeline, the uncalibrated raw data and rate files were processed with default parameters, respectively. Apart from the outlier detection step of Spec3, bad pixels were also identified and corrected through iterative sigma-filtering and a two-dimensional Gaussian kernel with the VIP package before Spec3. In Spec3, a Gaussian fit on the weighted average image of each band was used to determine the centroid for aperture photometry. A 2-FWHM (full width at half maximum) aperture was considered for spectrum extraction, and the background level was estimated from a surrounding annulus. 
    
        \begin{figure}
            \centering
            \includegraphics[width=\linewidth]{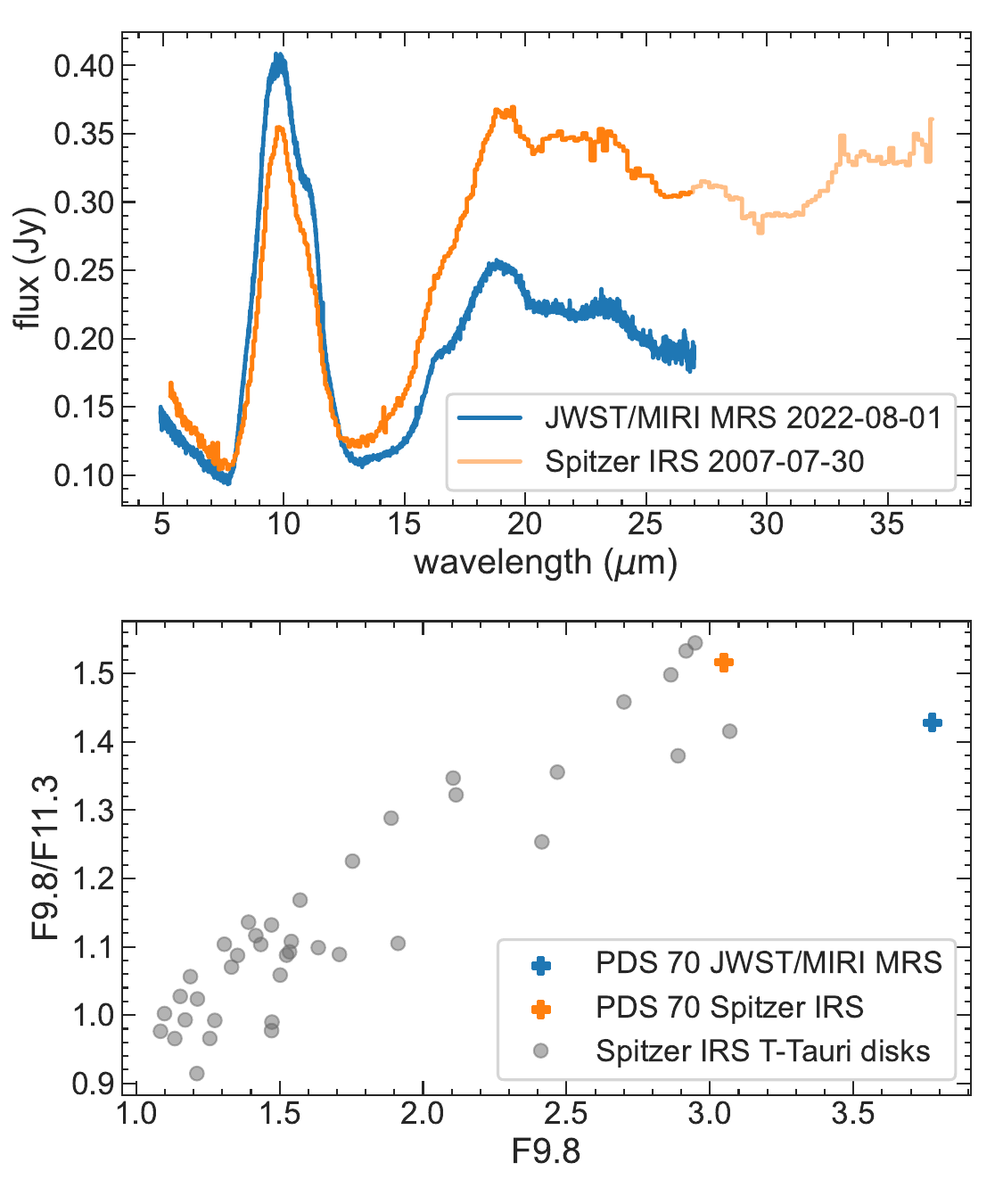}
            \caption{The MIRI and Spitzer spectra of PDS~70 and their shapes of 10 $\mu$m silicate bands. \textit{Upper}: MIRI spectrum is in blue, and Spitzer spectrum is in orange. Light orange shows full wavelength range of the Spitzer spectrum. Around $10~\mu$m, the MIRI spectrum has a higher flux and broader silicate band while the Spitzer spectrum shows higher flux levels beyond $12~\mu$m. \textit{Lower}: Band strength at 9.8 $\mu$m and the shape of 10 $\mu$m silicate band. Gray dots are disks in CASSIS database with low resolution spectra. Blue cross is the MIRI spectrum while orange cross is the Spitzer spectrum.}
            \label{fig:MIRI_Spitzer}
        \end{figure}

    \subsection{Characterization of the 10 $\mu$m silicate band}
    \paragraph{}
    Around 10 $\mu$m, MIRI spectrum shows a higher and broader flux emission compared to Spitzer while the fluxes are lower in the MIRI spectrum beyond 12 $\mu$m as shown in \fg{MIRI_Spitzer}. Calibration uncertainty of the MIRI spectrum is ruled out for any differences greater than 5~\% \citep{Perotti_etal2023}. The difference is  $\sim$10~\% below 12~$\mu$m and $\sim$40~\% above 12~$\mu$m. The reported spectrophotometric accuracy is 2–10~\% for Spitzer IRS \citep{Furlan_etal2006, Watson_etal2009} and 5.6\% ± 0.7\% for JWST/MIRI MRS \citep{Argyriou_etal2023}. The JWST Science Calibration pipeline version 1.12.5 that we used takes into account the time-dependent sensitivity of the MIRI MRS. This variability in the PDS~70 disk is not only found at mid-infrared wavelengths but also at shorter wavelength ranges \citep{Gaidos_etal2024}. At the longer wavelengths beyond 12 $\mu$m, we can also see clear forsterite bands at 16 and 19 $\mu$m in MIRI and Spitzer spectra. However, the presence of the 23 $\mu$m band is not obvious in the Spitzer spectrum, but we identify it in the MIRI spectrum.
    
    The strength of the 10 $\mu$m silicate band in the PDS~70 disk is unusually strong for a T-Tauri disk. To illustrate this, we measured the band strengths and shapes of the 10 $\mu$m silicate band for both MIRI and Spitzer spectra (following \citealt{vanBoekel_etal2003} and \citealt{Olofsson_etal2009}) and compared with a sample of T-Tauri disks and pre-transitional disks from the Spitzer CASSIS database \citep{Lebouteiller_etal2011} in the lower panel of \fg{MIRI_Spitzer}. The band strength at 9.8 $\mu$m (F9.8) is measured as
        \begin{equation}
            \text{F}9.8 = 1 + (f_{9.8\;\mu\rm m, cs} / <f_{\rm c}>),
        \end{equation}
    where $f_{9.8\;\mu\rm m,cs}$ is the linear continuum subtracted flux at $9.8$ $\mu$m, and $<f_{\rm c}>$ is the mean of the linear continuum. The linear continuum is drawn from 7.5 $\mu$m to 12.7 $\mu$m. We also measured the ratio of band strengths at 9.8 $\mu$m and 11.3 $\mu$m (F9.8/F11.3), representing the shapes of 10 $\mu$m silicate band. We can see that the PDS~70 disk has very strong 9.8 $\mu$m strength and convex shape, representing dominant amorphous small grain population in the disk \citep{vanBoekel_etal2003,Olofsson_etal2009}. 
    
    Disks with larger grains move to the lower left of the diagram in the lower panel of \fg{MIRI_Spitzer} \citep{vanBoekel_etal2003, Furlan_etal2006}. Clearly, the difference between the MIRI and Spitzer spectra does not follow this trend. The MIRI spectrum has smaller F9.8/F11.3 value, so we expect to see larger grains in the MIRI spectrum compared to Spitzer data (smaller F9.8 value). However, the F9.8 value in the MIRI spectrum is larger than in the Spitzer spectrum, indicating the presence of smaller grains. Another factor affecting the F9.8/F11.3 value is crystallinity, so the small grains might be highly crystalline in the PDS~70 disk. This behavior still does not align with the overall trend observed in other T-Tauri disks as shown in \fg{MIRI_Spitzer}. This, combined with the unusually strong silicate band, indicates that the grain population probed by these spectra may be different from that seen in other disks.

\section{Dust fitting method}
\label{sec:dustfittingtool}
\paragraph{}
We used the Dust Continuum Kit (DuCK; \cite{Kaeufer_etal2024}) to identify dust species in the PDS~70 disk. DuCK is a part of Dust Continuum Kit with Line emission from Gas (DuCkLinG), which is able to simultaneously account for the gas and dust contributions to mid-infrared spectra. However, we only use the dust continuum part, which is identical in its setup to the two-layer disk model introduced by \cite{Juhasz_etal2009, Juhasz_etal2010}. The model and fitting procedure are described in \sect{DuCKmodel}. \Sect{modelsetup} describes the modeled stellar photospheric spectrum for the PDS~70, the reduced MIRI spectrum onto a wavelength grid of lower and constant spectral resolving power, opacities of the dust from different grain shape models and laboratory measurement, and the temperature setup in DuCK model.

\subsection{DuCK model}
\label{sec:DuCKmodel}
\paragraph{}
DuCK is a retrieval model to fit the dust continuum of a protoplanetary disk spectrum \citep{Kaeufer_etal2024}. DuCK consists of the star, inner rim, optically thin disk surface, and optically thick midplane, and superposes these components \citep{Juhasz_etal2009, Juhasz_etal2010}. Thus, the modeled flux density is 
    \begin{equation}
        F_{\nu} = F_{\nu }^{\rm{star}} + F_{\nu}^{\rm{rim}} + F_{\nu }^{\rm{mid}} + F_{\nu }^{\rm{sur}},
    \end{equation}
where $F_{\nu }^{\rm{star}}$, $F^{\rm{rim}}_{\nu }$, $F^{\rm{mid}}_{\nu }$, and $F^{\rm{sur}}_{\nu }$ are the fluxes of the star, the inner rim, the midplane, and the disk surface, respectively. 
The flux of the inner rim ($F^{\rm rim}_\nu$) describes the emission coming from the inner edge of the disk and is described by the Planck function at the temperature of the inner rim ($T_{\rm rim}$) and a scaling factor (sc$_{\rm rim}$). The scaling factor represents the size of the emission area of the inner rim. The midplane flux summarizes the optically thick dust emission in a disk, that can be approximated by a radial temperature powerlaw with the exponent ($q_{\rm mid}$) between the minimum ($T_{\rm mid, min}$) and maximum temperatures ($T_{\rm mid, max}$) and another scaling factor (sc$_{\rm mid}$). The last component contributing to the model flux is the disk surface, which accounts for the optically thin dust emission from a disk. The temperature is following, analogous to the midplane, a radial temperature powerlaw with the exponent ($q_{\rm sur}$) between the minimum ($T_{\rm sur, min}$) and maximum temperatures ($T_{\rm sur, max}$). 

The fit quality of a model is determined by the likelihood using the model fluxes and the observational fluxes with their uncertainties for all spectral points \citep[see Eq. 14 in][]{Kaeufer_etal2024}. Moreover, the temperature parameters ($T_{\rm rim}$, $T_{\rm mid, min}$, $T_{\rm mid, max}$, $T_{\rm sur}$, and $q_{\rm mid}$) are sampled in a Bayesian way using MultiNest \citep{Feroz_Hobson2008, Feroz_etal2009, Feroz_etal2019} while the scale parameters and dust abundances are determined by using a non-negative least square fitting with SciPy \citep{Virtanen_etal2020}. Linear fitting during the Bayesian analysis decreases the number of parameters sampled by MultiNest and, therefore, decreases the computational cost of the fitting. \cite{Kaeufer_etal2024} shows that the retrieved results a largely consistent with a full Bayesian analysis of all parameters.

    \subsection{Fitting model setup}
    \label{sec:modelsetup}
    \paragraph{}
    The emission of the star ($F_{\nu }^{\rm{star}}$) is described by a modeled stellar photospheric spectrum, using PHOENIX code \citep{Baron_etal2010}, of an effective temperature $T_{\rm eff} = 4,000$ K, surface gravity $\ln(g) = 4.5$, and solar metallicity, used in \cite{Perotti_etal2023}. The modeled stellar spectrum was extrapolated ($\propto \lambda^{-2}$) to the mid-infrared wavelengths and was averaged every 15 data points to avoid biased fitting results at short wavelength due to high resolution of the stellar model. 
    
    We rebinned the MIRI spectrum ($R \sim 3000$) onto a wavelength grid of constant spectral resolving power ($R \sim 500$). We first generate a wavelength grid that has a constant spectral resolution. From the spectrum, we measured the mean and standard deviation of fluxes within every wavelength grid. Thus, we could find the normal error distribution of the equally weighted spectrum. This method is valid because we do not fit gas and dust emissions together and assume the spectrum is fully due to dust emission. \cite{Perotti_etal2023} only identify H$_2$O and CO$_2$ emission lines, and the spectrum is not dominated by these molecular emissions. The typical errorbar of individual spectral points of the MIRI spectrum is reported as 0.2 mJy in \cite{Perotti_etal2023}, but we have errorbars as large as $1-3$ mJy because of the weak gas emission. We note that the fitting results are not sensitive to the value of spectral resolving power whether it is $R \sim$ 300 or 700, and $R \sim 500$ is good enough to represent dust bands present in the spectrum.

    We use three different sets of opacities to fit the MIRI and Spitzer spectra: Distribution of Hollow Sphere (DHS, \cite{Min_etal2005}), Gaussian Random Field (GRF, \cite{Min_etal2007}), and optical data obtained from laboratory-based aerosol spectroscopy (hereafter Aerosol) \citep{Tamanai_etal2006, Tamanai_etal2009}. The optical constants, which are derived from experimentally measured reflectance spectra by using the Kramers–Kronig relations (e.g., \citealt{Jaeger_etal1998}), are applied for both GRF and DHS models and listed in Table \ref{tab:optical_const}. These optical constants are converted to opacities using two grain shape models of irregularly shaped dust grains, GRF and DHS. On the other hand, the extinction spectra can be obtained via Aerosol spectroscopy. Particulates are suspended and retained in nitrogen (N$_2$) gas flow during the in-situ measurements. Since the refractive index of the gas is comparable to that of vacuum, these experimentally obtained extinction spectra are not affected by the medium at all. 

    Table \ref{tab:optical_const} shows the optical constants of amorphous and crystalline silicates as well as amorphous SiO$_2$ for GRF and DHS calculations. We select two types of crystalline silicates: forsterite and enstatite, together with stoichiometric amorphous Mg$_2$SiO$_4$, Mg$_2$SiO$_4$, and SiO$_2$. In DHS, the inner volume fraction of hollow sphere ($f_{\rm max}$) was set to be 0.99 for crystalline dust and 0.7 for amorphous dust \citep{Min_etal2005,Min_etal2007}. All dust grains have sizes of 0.1, 1, 2, 3, 4, 5 $\mu$m. Grain sizes greater than 5 $\mu$m are not considered in DuCK model because the band strength becomes very weak already for 5 $\mu$m-sized grains. We note that DuCK does not use grain size distributions, so the absence of intermediate grain sizes between the detected smallest and largest sizes in results does not imply that these sizes do not exist in the disk. Rather, the best-fit DuCK result is consistent with a spectrum, therefore no further grain sizes are necessary. In Aerosol, we pick out the individual particle size around 1 $\mu$m for each measurement; however, the extinction spectra contain the agglomerates composed of micron-sized particles as well. Absorption efficiencies of amorphous and crystalline Mg$_2$SiO$_4$ in DHS and GRF together with the Aerosol spectrum are shown in \fg{Qcompare}. We note that Aerosol spectra are experimentally measured non-quantitative absorbance, so the intensity varies depending the particle concentration. Therefore, we normalized the aerosol spectra using the right y-axis in \fg{Qcompare}. Moreover, the DHS and GRF opacities are converted to absorption efficiencies by multiplying them by $4/3 \pi a \rho \kappa_{\rm abs}$, where $a$ is the grain size, $\rho$ is the material density, and $\kappa_{\rm abs}$ is the DHS and GRF opacities. In the case of amorphous Mg$_2$SiO$_4$ (upper panel), overall band profiles of absorption efficiencies calculated by the GRF and DHS are well consistent with the Aerosol. By contrast, absorption peaks of crystalline Mg$_2$SiO$_4$ calculated by the DHS undergo a clear redshift at longer wavelengths ($\lambda >$ 15 $\mu$m) compared to the GRF and Aerosol spectra (bottom panel). First, we focus on Mg-rich silicates to fit the spectrum, but we also try to include Fe-rich silicates, such as fayalite (crystalline Fe$_2$SiO$_4$) and amorphous MgFeSiO$_4$ and MgFeSi$_2$O$_6$ in \sect{fittingFe}.

    \begin{figure}
    \centering
    \includegraphics[width=\linewidth]{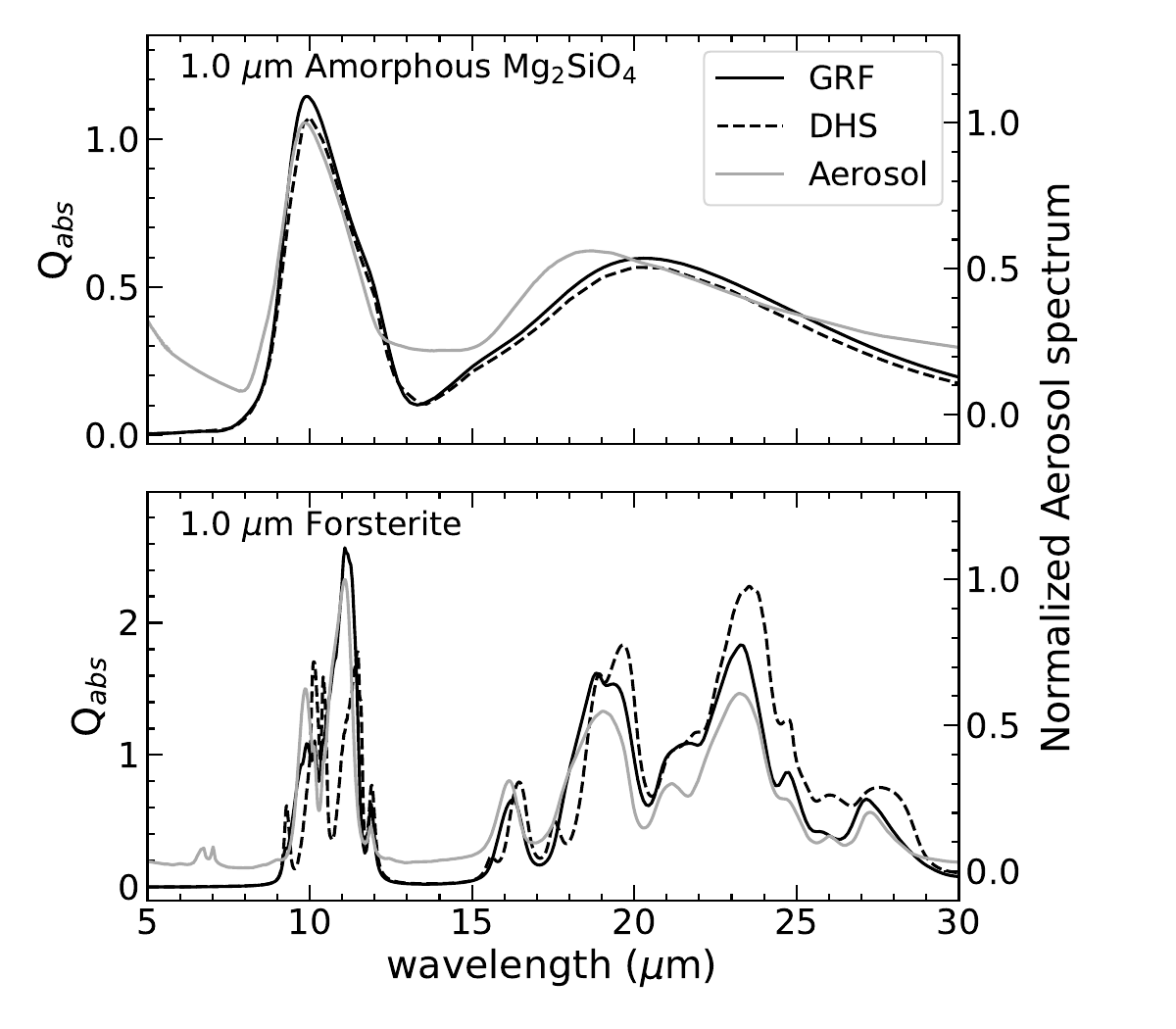}
    \caption{Comparison among absorption efficiencies for amorphous Mg$_2$SiO$_4$ (upper panel) and forsterite (lower panel). GRF (black solid line) and DHS (dashed line) are plotted for $Q_{\rm abs}$ on the left y-axis, and Aerosol spectra (gray solid line) are normalized on the right y-axis.}
    \label{fig:Qcompare}
    \end{figure}

    In DuCK, we set the priors for temperatures of the inner rim, the disk surface, and the midplane as $T_{\rm rim} = [1000,1500]$, $T_{\rm sur} = [150,700]$, $T_{\rm mid, min} = [10, 1000]$, $T_{\rm mid, max} = [10,1000]$. The exponent for the temperature gradient of the midplane is $q_{\rm mid, min} = [-1.5,-0.1]$. We adjusted the priors to be large enough to cover full posterior probability distributions within a reasonable temperature range for a protoplanetary disk. The posterior probability distributions for the best model in this study (discussed in the next section) is shown in \fg{GRFposterior}. In addition, DuCK allows for measuring disk surface flux with a single temperature instead of the temperature gradient.

\section{Dust fitting with different opacities}
\label{sec:results}
\paragraph{}
We first conducted a fitting process using a temperature gradient for the disk surface, which results in $T_{\rm sur, min} = 460$ K and $T_{\rm sur, max} = 490$ K. Due to the small temperature range, fitting can be performed with a single temperature for the disk surface. Importantly, the results remain consistent with the single disk surface temperature. Thus, we proceed to fit the PDS~70 spectrum using a single temperature for the disk surface. Mass fractions of each species and grain sizes of the fitting with GRF, DHS, and Aerosol are summarized in Table \ref{tab:results}, and temperatures of the inner rim, the disk surface, and the midplane are plotted in \fg{temperatures}. Fitting results are also shown in \fg{GRFfitting}, \fg{DHSfitting}, and \fg{Aerosolfitting}.

\begin{figure}
\centering
\includegraphics[width=\linewidth]{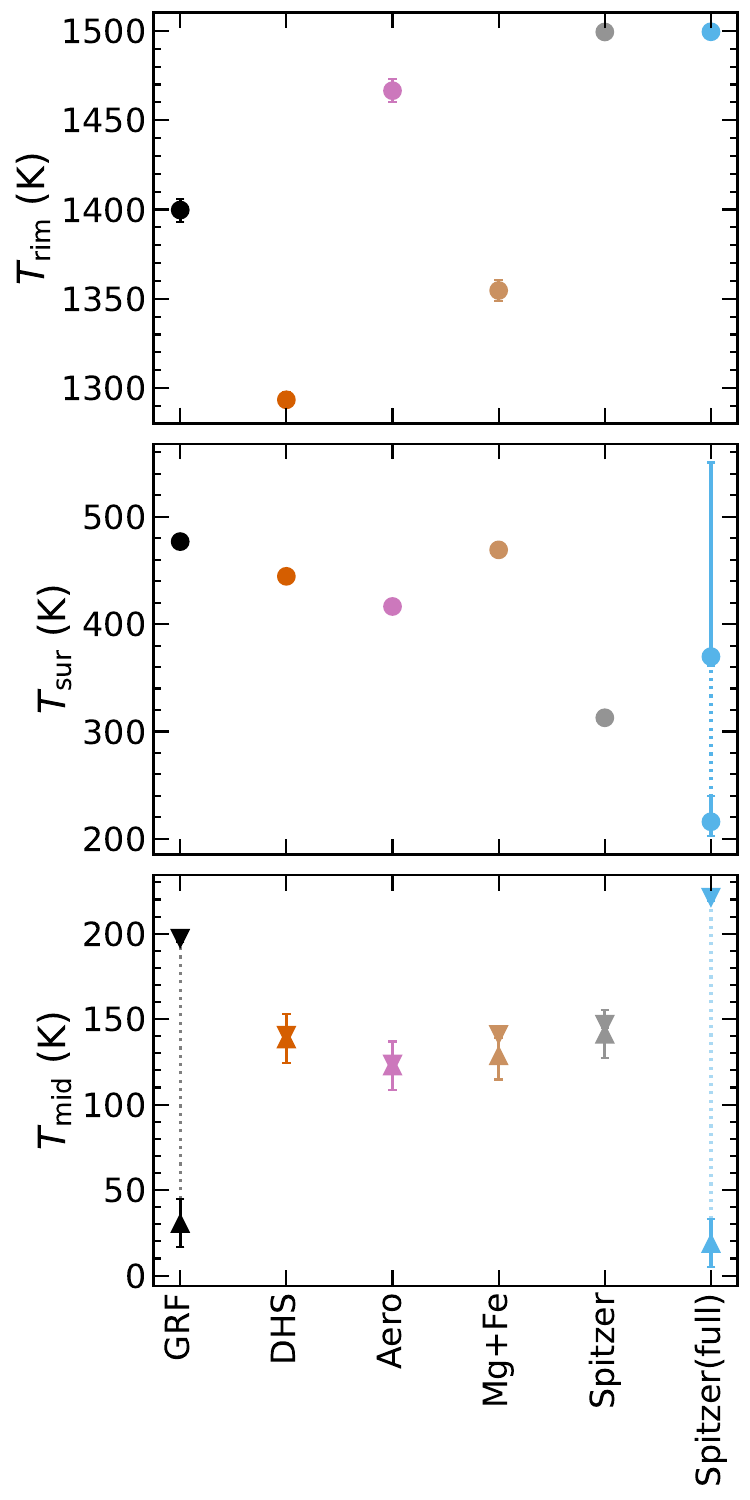}
\caption{Temperatures of dust in the inner rim (top), disk surface (middle), and midplane (bottom) of dust fitting models in Table \ref{tab:results}. For Spitzer(full), the dotted line in the temperature of the disk surface indicates the temperature gradient across the disk surface; the other models assume a single temperature.}
\label{fig:temperatures}
\end{figure}

\begin{table*}[]
\centering
\begin{threeparttable}
\caption{Optical constants used to generate opacities with GRF and DHS.  }
\label{tab:optical_const}
\begin{tabular}{ll}
\hline\hline
 species & reference \\ \hline
  Forsterite & \cite{Servoin_Piriou1973} \\
 Enstatite & \cite{Jaeger_etal1998} \\
 Fayalite &  \cite{Fabian_etal2001}\\
 Amorphous Mg$_2$SiO$_4$ & \cite{Henning_Stognienko1996}\tnote{1}\\
 Amorphous MgSiO$_3$ & \cite{Dorschner_etal1995} \\
 Amorphous SiO$_2$ & \cite{Henning_Mutschke1997} \\
 Amorphous MgFeSiO$_4$ & \cite{Dorschner_etal1995} \\
 Amorphous MgFeSi$_2$O$_6$ & \cite{Dorschner_etal1995} \\
\hline
\end{tabular}
\begin{tablenotes}
    \item[1] The optical constant can be found from this website: \url{https://www2.mpia-hd.mpg.de/homes/semenov/Opacities/RI/new_ri.html}
  \end{tablenotes}
\end{threeparttable}
\end{table*}

\textit{Model GRF}: Using GRF opacities, the dominant dust species is amorphous silicates contributing $\sim$79 \% of the surface layer dust mass, with  crystalline silicate at $\sim$20 \%. We find no evidence for enstatite and amorphous SiO$_2$. Most silicate species are a few microns in size while amorphous MgSiO$_3$ shows significant mass fraction of 0.1 $\mu$m grains with $\sim$33 \%. The inner rim temperature is $\sim$1400 K, the disk surface is $\sim$500 K, and the midplane has temperature gradient from 25 K to 200 K as shown in \fg{temperatures}. The mean residual of this fit is around 1.3 \%, and the overall residual is shown in \fg{GRFfitting}. The largest residual appears at 8 $\mu$m because the shapes of GRF dust opacities do not perfectly match at the start of the slope of 10 $\mu$m silicate band. Other than the spike, the overall quality of the fit is in good agreement with the MIRI spectrum.

    \begin{figure*}
    \centering
    \includegraphics[width=\linewidth]{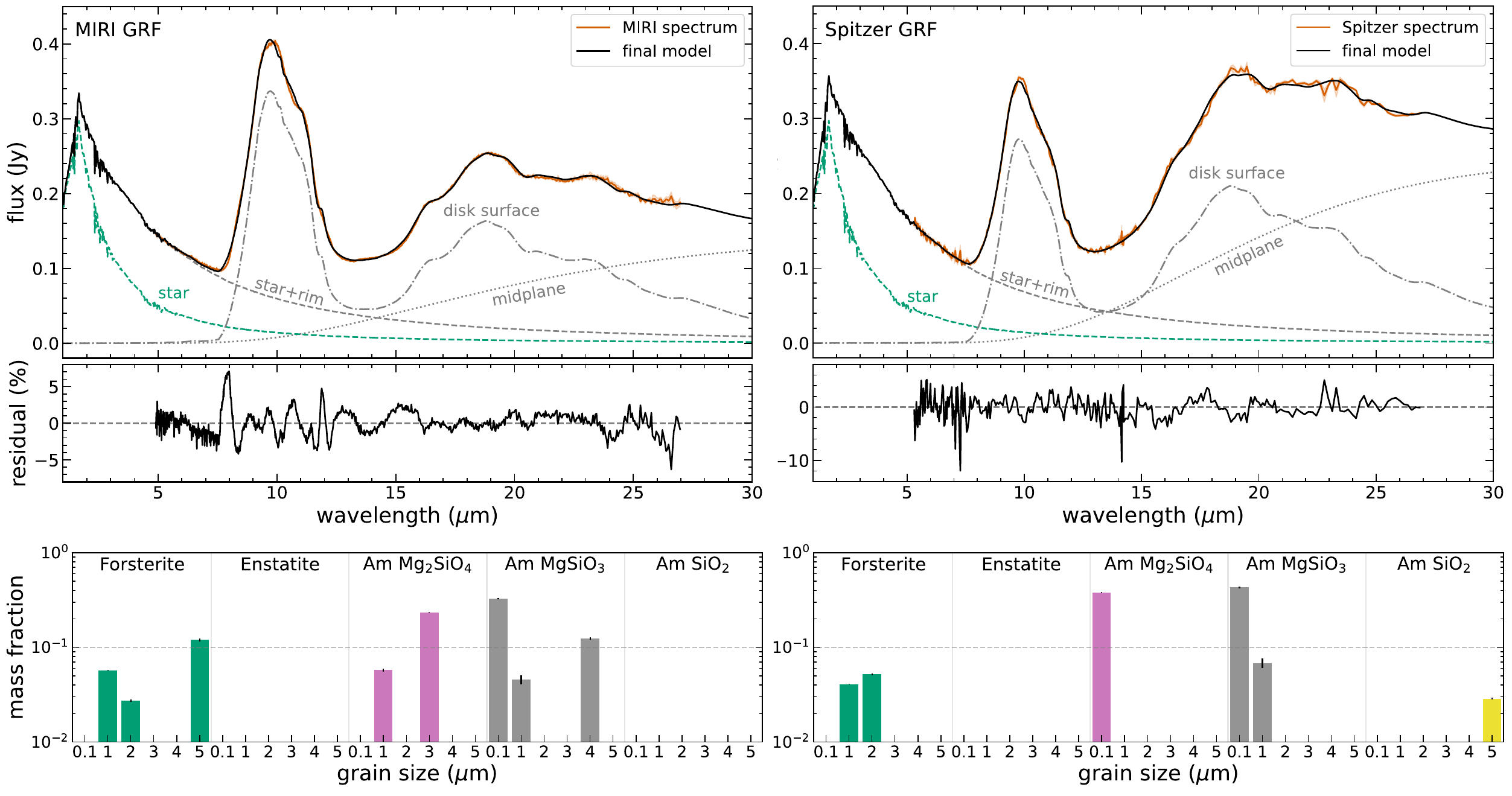}
    \caption{Fitting with GRF opacity. The modeled spectrum matches well with the MIRI spectrum with the mean residual less than 2 \%. Amorphous silicates dominate the dust mass abundance, and 0.1~$\mu$m-sized amorphous MgSiO$_3$ is the most abundant species. Enstatite and amorphous SiO$_2$ do not show up from the fitting.}
    \label{fig:GRFfitting}
    \end{figure*}

\textit{Model DHS}: Using DHS opacities, amorphous silicate dominate the spectrum, even more than in Model GRF at $\sim$94 \% of the optically thin dust mass fraction, with forsterite at $\sim$5 \%, i.e. a significantly lower crystallinity compared to Model GRF. 0.1 $\mu$m-sized amorphous MgSiO$_3$ is still the most dominant dust grain species, and 0.1 $\mu$m-sized forsterite is found at a mass fraction of $\sim$3 \%. Unlike Model GRF, grain sizes larger than 2 $\mu$m are not found in this model. Enstatite and amorphous SiO$_2$ are still not detected. The inner rim temperature is around 1300 K, which is lower than Model GRF. The disk surface temperature is $\sim$450 K, similar to Model GRF, while the midplane temperature is $\sim$140 K, much narrower temperature range as shown in the lower panel of \fg{temperatures}. The mean residual is  1.5 \%, as shown in \fg{DHSfitting}. Around 20 $\mu$m, the forsterite band of our model is slightly red-shifted and overestimated compared to the MIRI spectrum while Model GRF does not show this mismatch. This red-shifted forsterite band on DHS is also seen in opacities of forsterite in \fg{Qcompare}, compared to GRF.

\textit{Model Aerosol}: The overall shape of the fitting aligns with the spectrum as shown in \fg{Aerosolfitting}. Nevertheless, the peak of 10 $\mu$m silicate band and 21 $\mu$m band are overestimated while the 16 $\mu$m band is underestimated. Both shoulders of the 10 $\mu$m silicate band are also underestimated or overestimated. As other models with GRF and DHS, amorphous silicates dominate the spectrum with 95 \%. However, amorphous Mg$_2$SiO$_4$ is the dominant species among amorphous silicates. Unlike Model GRF and DHS, enstatite and amorphous SiO$_2$ are found with $\sim$2 \% and $\sim$7 \%, respectively. The inner rim temperature is relatively higher than other models at $\sim$1450 K, but the disk surface temperature is slightly lower at $\sim$400 K. The midplane temperature is $\sim$125 K, similar to Model DHS. We find that the use of Aerosol absorption efficiencies results in fitting residuals that are significantly larger than for the other two models. The mean residual is around 3.7 \%, and the maximum residual is around 15 \% as shown in \fg{Aerosolfitting}. This is expected, because the Aerosol measurements are limited in the range of grain sizes and shapes that are sampled.

We use the logarithm of the Bayes factor to assess the quality of the fits presented above. The bayes factor is defined as the difference between the Nested sampling Global Log-Evidence of models ($\ln({\rm Model})$), denoted as $\ln B_{1, 2} = \ln({\rm Model1}) - \ln({\rm Model2})$. The Nested sampling Global Log-Evidence comes as a result of MultiNest. If the bayes factor for Model1 and Model2 ($\ln B_{1, 2}$) is smaller than 1, it represents that Model1 has no evidence over Model2 \citep{Trotta2008}. Model DHS and Model Aerosol have negative Nested sampling Global Log-Evidence, while Model GRF is $\ln({\rm Model\; GRF})=937$. Thus, $\ln B_{\rm GRF,~DHS}$ and $\ln B_{\rm GRF,~Aerosol}$ are very negative, and Model GRF is clearly the preferred model over others. 
Therefore, we focus mainly on the GRF opacity in the remainder of this study.

Model GRF and Model DHS show the forsterite mass fractions of 20.4\% and 4.6\%, respectively. In Model GRF, the dominant forsterite mass comes from the largest grains ($5 \mu$m), which contribute the least to the flux compared to the smaller grains. To test the dependence of the crystalline mass fraction on the largest grain size, we fitted the same GRF model but excluded the large forsterite grains in Model GRF4$\mu$m (forsterite up to 4 $\mu$m) and Model GRF2$\mu$m (forsterite up to 2 $\mu$m). As the large grains are depleted, the crystalline mass fraction decreases. In Model GRF4$\mu$m and Model GRF2$\mu$m, the mass fractions of the largest grains slightly decrease to $\sim$18 \% and $\sim$12. \%, respectively. Although we neglect the largest forsterite grains in Model GRF4$\mu$m and Model GRF2$\mu$m, crystallinity remains at the same magnitude, and the largest forsterite is always preferred in those models. Moreover, the mass fractions for each grain size also depend on the grain shape model. Model DHS has no contribution to crystallinity from the large grains because forsterite bands quickly decrease in strength for 1 $\mu$m-sized grains. Thus, the crystallinity in mass fraction may depend on the largest grain size given in the model and grain shape model, but the crystallinity is always greater than 10 \% based on Model GRF, Model GRF4$\mu$m, and Model GRF2$\mu$m.

\section{Iron-containing silicates}
\label{sec:ironSilicate}
\paragraph{}
In the previous section, we used opacities from pure Mg silicates, but iron, being highly refractory, condenses into the solid phase. Iron-containing silicates, another end member of olivine and pyroxene stoichiometry, could exist in the PDS~70 inner disk. Iron-containing silicates have been detected in warm debris disks \citep{Olofsson_etal2012}, and in a small number of gas-rich T-Tauri disks \citep{Olofsson_etal2010}. We explore the iron content in the PDS~70 disk through two methods: measuring the peak of the 23-24 $\mu$m crystalline silicate band (\sect{peakposition}) and dust fitting with Fe-rich silicates (\sect{fittingFe}). We note that iron may also exist as FeS or metallic iron. While metallic iron contributes opacity, it lacks infrared vibrational resonances, making its direct detection difficult with our method. FeS also has generally weak mid-infrared resonances. Its main band centered at 23.5 $\mu$m is too broad in the region where the 18 $\mu$m amorphous silicate band dominates the spectrum. Thus, we have not included this material in our study.

    \subsection{The 23-24 $\mu$m dust band}
    \label{sec:peakposition}
    \paragraph{}
    In crystalline olivine, although the absorption peaks are mostly contributed by Si–O–Si stretching and O–Si–O bending vibrations up to around 20 $\mu$m, translational motions of metal cations within the oxygen cage and complex translations involving metal and Si atoms lead to complex bands beyond 21.28 $\mu$m (more details in e.g., \citealt{Servoin_Piriou1973, Dorschner_Henning1986, Hofmeister1997,Henning_etal2005}). Iron plays an important role in the wavelength region between 23 and 24 $\mu$m (e.g., \citealt{Jaeger_etal1998, Koike_etal2003, Tamanai_Mutschke2010}). As the amount of iron increases in olivine, this 23-24 $\mu$m band undergoes a redshift up to the iron content of 40 \%. However, the intensity of this peak weakens beyond this point. When the iron content exceeds 60 \%, this peak is no longer visible \citep{Koike_etal2003}. For instance, \cite{Tamanai_Mutschke2010} showed that forsterite has a peak at 23.27 $\mu$m whereas this peak is located at 23.91 $\mu$m for 20 \% iron content of olivine based on Aerosol spectra. Thus, the position of this peak gives us a clue to the presence of iron in olivine dust grains.

    We pay attention to the peak around 23-24 $\mu$m of the MIRI spectrum. By fitting a 1D Gaussian curve, the peak position is located at 23.44 $\mu$m. Likewise, we also derive the peak from the Spitzer data which is found to be 23.47$\pm$0.01 $\mu$m. The fitted 23-24 $\mu$m band and the peak positions of the full spectral resolution MIRI spectrum and rebinned MIRI spectrum (R=500) are shown in \fg{23micron_peak}. The rebinned spectrum is the one used for dust fitting. These results correspond to a 10 \% iron content olivine, which has its peak situated at 23.46 $\mu$m \citep{Tamanai_Mutschke2010}. We note that the error on the measured wavelength does not include systematic errors resulting from slightly different choices of the underlying continuum.

    \subsection{Dust fitting with Fe-rich silicate dust}
    \label{sec:fittingFe}
    \paragraph{}
    In addition to Model GRF, we also fitted the MIRI spectrum with Fe-rich silicate species, including fayalite, amorphous MgFeSiO$_4$, and amorphous MgFeSi$_3$O$_6$, and we refer to this as Model Mg+Fe. The model agrees well with the MIRI spectrum, and the mean residual is 1.1 \%, as shown in \fg{GRFall}. Within this model, enstatite and fayalite are found to be less than 1 \% in mass fraction, and we exclude those dust species from the figure. The model fit retrieves large Fe-rich amorphous silicate grains, while pure Mg amorphous silicates are more favored for small grains. 
    
    Large amorphous silicates have broad and weak band resonances and weak flux contributions compared to small amorphous and crystalline silicates, but the mass fraction is proportional to the grain size retrieved. Thus, small flux contributions from the large grains can result in large mass contributions. However, we did not find a satisfactory fit to the MIRI spectrum using only Fe-rich silicates, i.e. excluding pure Mg silicates.
    These results demonstrate that pure Mg silicates are required to fit the spectrum and dominate the PDS~70 disk, while the amount of Fe-rich silicate is still ambiguous. 

    \begin{figure}
        \centering
        \includegraphics[width=\linewidth]{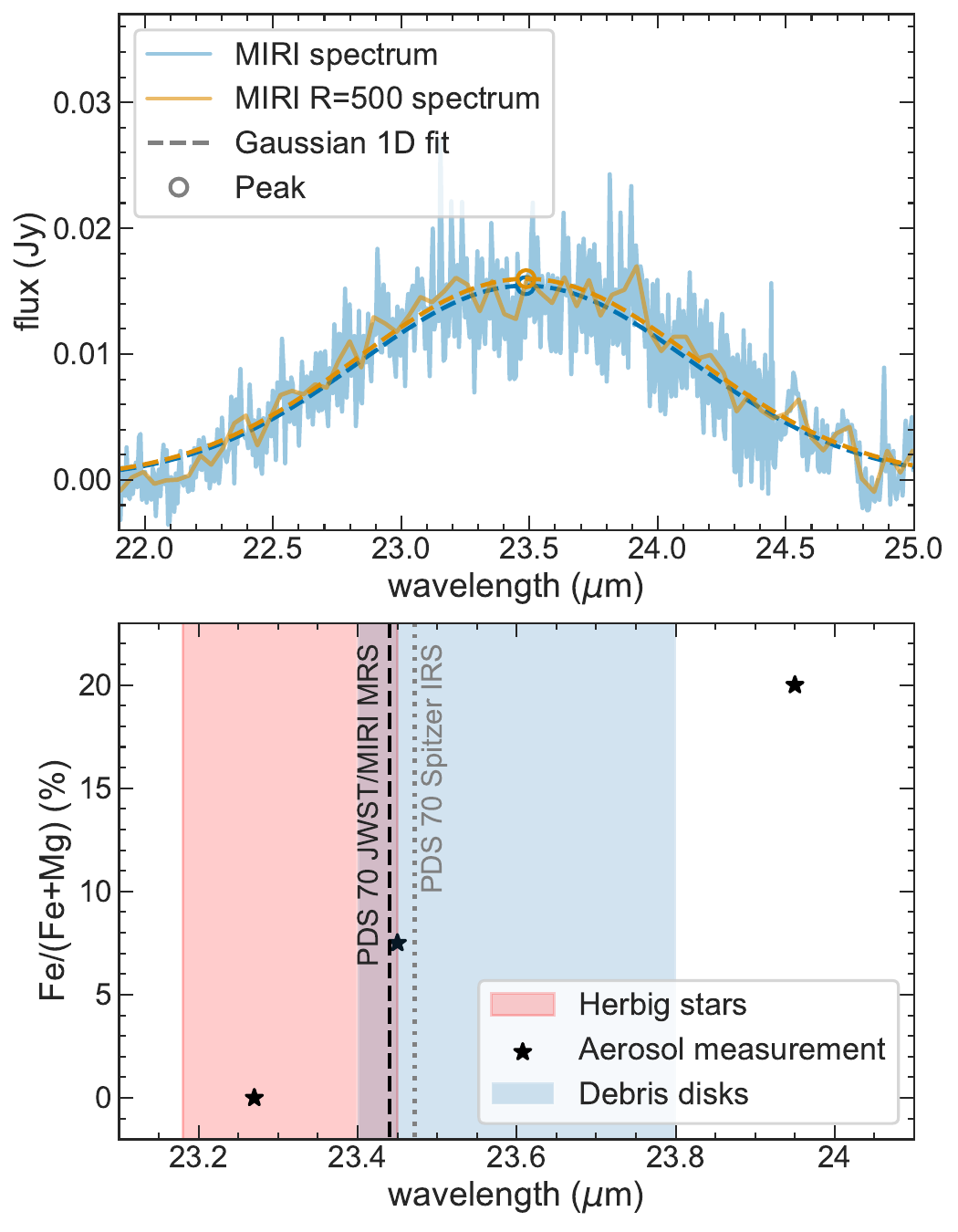}
        \caption{Measurement of the 23-24 $\mu$m band peak. The MIRI spectrum was fitted with a 1D gaussian curve (blue dashed line) and found the peak at 23.4 $\mu$m (blue circle) in the upper panel. MIRI spectrum with R=500 is also plotted and fitted with the 1 D gaussian curve (yellow solid and dashed lines), and the gaussian curve also peaks at 23.4 $\mu$m (yellow circle). Lower panel shows the locations of the peaks for the MIRI and Spitzer spectra together with Aerosol measurements. Regions where the peaks locate for Herbig Ae/Be stars and debris disks are shaded in red and blue, respectively.}
        \label{fig:23micron_peak}
    \end{figure}

    \begin{figure}
        \centering
        \includegraphics[width=\linewidth]{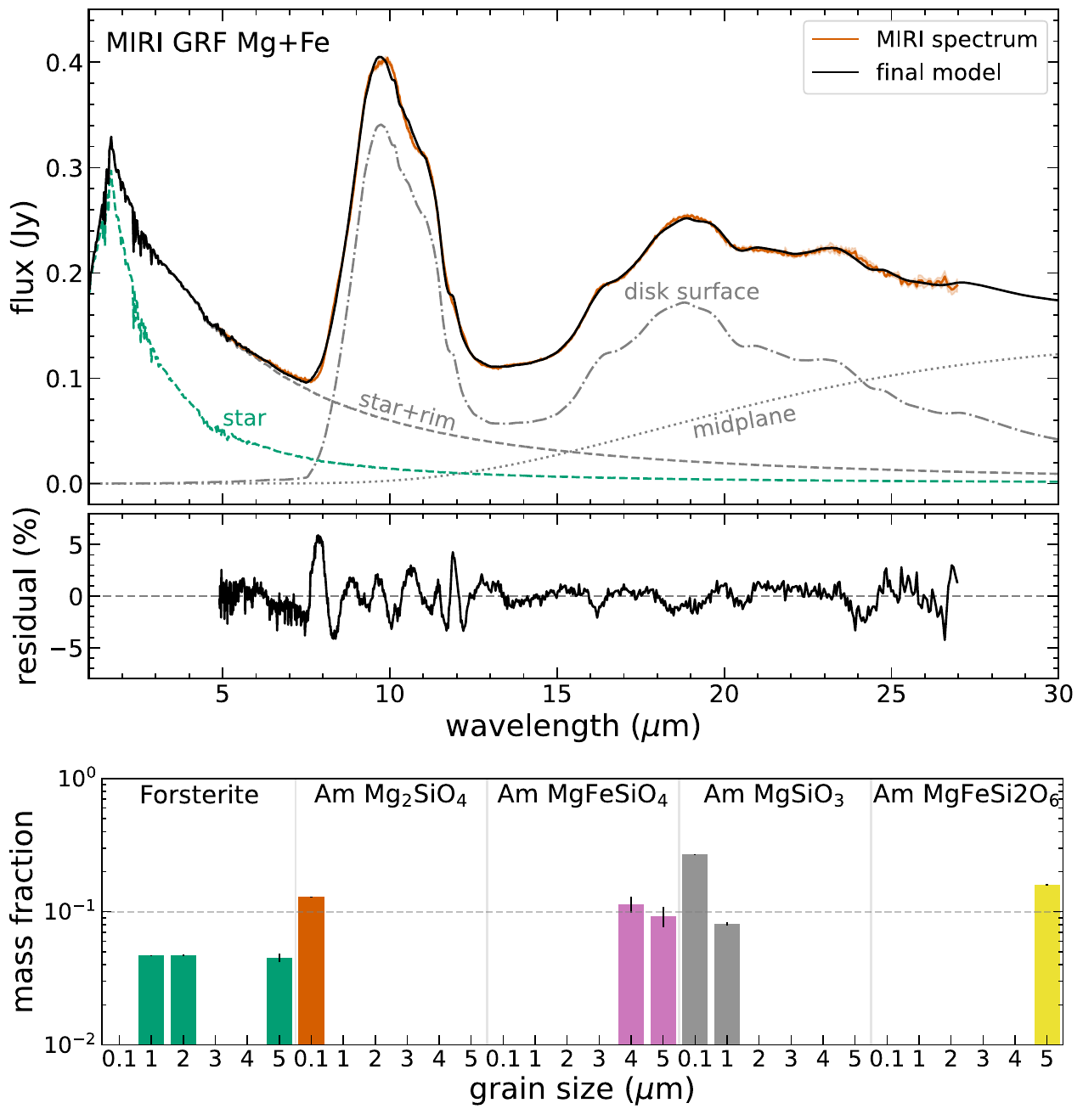}
        \caption{Fitting with additional Fe-rich dust GRF opacities from Model GRF. Overall fitting result is as good as Model GRF, especially beyond 14 $\mu$m. Mass fractions are only shown for dust species detected greater than 1 \%. }
        \label{fig:GRFall}
    \end{figure}

\section{Dust fitting for Spitzer data}
\label{sec:fittingSpitzer}
\paragraph{}
We also fitted the Spitzer spectrum with GRF opacities in two ways: 1) the same wavelength range with the MIRI spectrum (Model Spitzer), 2) the full wavelength range of Spitzer spectrum (Model Spitzer(full)). The results of Model Spitzer and Model Spitzer(full) are shown in the right panel of \fg{GRFfitting} and \fg{SpitzerFull}, respectivly. Because the Spitzer covers a broader wavelength range than MIRI, we allow the DuCK to constrain temperature gradient across the disk surface rather than assuming a single temperature in Model Spitzer(full). The temperatures are shown in \fg{temperatures}.

The inner rim temperatures derived from both Spitzer spectrum fittings are significantly higher than that obtained from the MIRI spectrum, reaching the upper limit, 1500~K, of our temperature priors in DuCK. We did not raise the upper limit because silicate dust sublimates above $\sim$1500 K \citep{Duschl_etal1996,Kama_etal2009}. We note that the precise inner rim temperature does not affect the analysis of dust mineralogy in the disk surface. The higher inner rim temperature derived from the Spitzer spectrum compared to the MIRI spectrum can be attributed to the differences in their flux at short wavelengths. The Spitzer spectrum has a higher flux around 5 $\mu$m, which is more sensitive to the hotter inner regions of the disk. In contrast to the inner rim, the disk surface temperature derived from the Spitzer spectrum is cooler than the MIRI spectrum because of the lower flux around the 10 $\mu$m silicate band, where warm dust dominates, and the increased flux at longer wavelength, where colder dust contributes. 

Both models indicate that amorphous silicates (Mg$_2$SiO$_4$ and MgSiO$_3$) with sizes ranging from 0.1 to 1 $\mu$m make up 85~\% of the total mass fraction in the disk surface. Crystalline silicates still do not show evidence of enstatite, but forsterite is clearly detected. Model Spitzer identifies only 1 and 2 $\mu$m forsterite grains with a mass fraction of $\sim$10~\% while Model Spitzer(full) finds 0.1, 1, 2, and 5 $\mu$m forsterite grains. Because of the large 5 $\mu$m sized grains, Model Spitzer(full) has a higher mass fraction of forsterite at $\sim$15~\% in total. The values are summarized in Table \ref{tab:results}.

Both MIRI and Spitzer spectra show a dominance of amorphous silicates (Mg$_2$SiO$_4$ and MgSiO$_3$), contributing $\sim$80 \% of the total mass fraction with a preference for SiO$_3$ stoichiometry. However, the model describing MIRI spectrum shows larger grain sizes, typically a few microns, compared to the model for Spitzer spectrum, which is dominated by 0.1~$\mu$m grains. In addition, the crystallinity derived from the MIRI spectrum ranges around 10-20 \%, depending on given largest grains, while Spitzer spectrum shows around 10-15 \% crystallinity.

    We also measured the mass of $0.1~\mu$m amorphous MgSiO$_3$ in the disk surface with
    \begin{equation}
        M_{\rm dust} = \dfrac{F(\nu)d^2}{B(\nu,T)\kappa(\nu)},
    \end{equation}
    where $F(\nu)$ is the flux from the amorphous MgSiO$_3$ in the disk surface, $d$ is the distance to PDS~70, $\kappa(\nu)$ is the opacity of $0.1 \mu$m amorphous MgSiO$_3$, $B(\nu,T)$ is the Planck function at frequency $\nu$, and the single disk surface temperature $T$. We find $4.5\times 10^{-7}$ M$_{\oplus}$, which is 32.8 \% of total dust mass in the inner disk surface, as the result of Model GRF, so the total dust mass in the inner disk surface is $1.37\times 10^{-6}$ M$_{\oplus}$. In Spitzer spectrum, amorphous MgSiO$_3$ is $4.8\times 10^{-7}$ M$_{\oplus}$, which is 37.6 \% of the total dust mass in the disk surface, so the total dust mass is $1.27\times 10^{-6}$ M$_{\oplus}$. The dust mass in the inner disk surface is similar between MIRI and Spitzer spectra. We note that only a small fraction of the total dust mass is in the disk surface, and \cite{Benisty_etal2021} infer the total dust mass within 18~au as $0.08-0.36$~M$_{\oplus}$.
    
\section{Discussion}
\label{sec:discussion}
\paragraph{}
Our fitting results agree on the dominance of small amorphous silicates and the lack of enstatite, and we found around or less than 10~\% iron content in crystalline silicates. Moreover, the variability of PDS~70 between JWST/MIRI MRS and Spitzer IRS can be seen in \fg{MIRI_Spitzer} and is also reported in \cite{Perotti_etal2023}. In this section, we discuss iron content in crystalline silicates in \sect{discuss_iron}, the absence of enstatite in \sect{discuss_enstatite}, the variability between MIRI and Spitzer data in \sect{discuss_variability}, and the origins of small silicate dust in \sect{discuss_origin}.

    \subsection{Iron content in crystalline silicates}
    \label{sec:discuss_iron}
    \paragraph{}
    We find that crystalline silicates in the PDS~70 disk contain up to $\sim$10 \% of iron based on the peak position of the 23-24 $\mu$m band. We note that in our theoretical grain shape models (e.g., GRF), the peak position shifts toward longer wavelengths with increasing grain size, as shown in \fg{23micron_shift}. That would imply that the Areosol-derived iron content - based on $\sim$ 1 $\mu$m sized grains - is in fact an upper limit. However, Aerosol experiments using samples of larger forsterite grains \citep{Tamanai_Mutschke2010} do not show a shift of the 23 $\mu$m band to longer wavelengths, i.e. different from the GRF grain model results. Taken together, we conclude that the iron content in the crystalline olivine is likely around or less than 10 \%. We note that we do not consider scattering opacities, which become important for optically thick dust clouds and large grains (a few microns). However, our analysis shows the dust observed in PDS~70 is dominated by optically thin, relatively small grains ($\sim$1~$\mu$m), so we only consider absorption coefficients in this study.
    
    It is interesting to compare the iron content in olivines in the PDS~70 disk to that found in gas-rich protoplanetary disks and debris disks. \cite{Juhasz_etal2010} and \cite{Olofsson_etal2010} found that Spitzer IRS spectra of Herbig Ae/Be and T-Tauri stars can be well fitted using forsterite, with only a handful of disks requiring iron-containing olivines. Limits on the iron content of less than 2 \% in olivine were found in the outer disks of the Herbig Ae/Be stars and T-Tauri stars, using Herschel detection of the 69 $\mu$ band \citep{Sturm_etal2013}. In warm debris disks like HD69830 and HD113766A, the presence of 20~\% iron-containing olivine has been reported in \cite{Tamanai_Mutschke2010, Olofsson_etal2012}. In contrast, \cite{devries2012} constrained the iron content in olivine in the cold debris disk of $\beta$ Pic to be 1 \%. These observations suggest that cold outer-disk olivines in gas-rich disks and debris disks are very iron-poor, while warm inner-disk olivines in gas-rich disks are generally very iron-poor but may contain some iron in warm debris disks, reaching the values up to 20 \%. Our limits on the iron content in olivines in the PDS~70 disk are consistent with those typical for the inner regions of gas-rich disks.

    \subsection{Lack of enstatite}
    \label{sec:discuss_enstatite}
    \paragraph{}
    Remarkably, we find no evidence for warm inner disk enstatite, but enstatite has been detected in many other inner disks. \cite{Bouwman_etal2008} investigated crystalline silicates in seven T-Tauri disks observed by Spitzer IRS and found that the inner warm dust at $\sim$1~au is more enstatite-rich, while the cold outer disk (around 5-15~au) is more rich in forsterite. \cite{Juhasz_etal2010} found that mid-infrared spectra of Herbig Ae/Be stars observed by Spitzer tend to show a higher abundance of enstatite than forsterite when analysing the shorter wavelength regions ($7-17~\mu$m), representing the inner disk, while it is vice versa in the longer wavelength region ($17-35~\mu$m), representing the outer disk \citep{Jang_etal2024}. \cite{Olofsson_etal2010} analyzed the dust composition of 58 T-Tauri disks observed with Spitzer IRS and found a similar fractions of forsterite and enstatite in the inner disk and a higher fraction of forsterite in the outer disk. Thus, in the inner gas-rich disks, enstatite is relatively abundant compared to the outer disk.

    Interestingly, \cite{Harker_etal2023} show that typical mass ratios of enstatite to forsterite in solar system comets observed by Spitzer are similar or greater than 1, but with large variations between individual comets. This could indicate that, in the comet forming region of the proto-solar nebula, significant reservoirs of both forsterite and enstatite were present. The building blocks of comets consist of small micron-sized grains, and the temperatures of comets are too low for parent-body processing. The Stardust mission that returned samples from the comet Wild 2 contained both forsterite and enstatite \citep{Brownlee2014}. All taken together, these findings show that small enstatite grains can be present in the icy regions of protoplanetary disks. 
    
    However, in most of our retrievals of MIRI and Spitzer spectra of PDS~70, enstatite is not detected except for Model Aerosol. Model Aerosol has limitations in grain sizes and is not well fitted with the spectrum. The high abundance of inner-disk enstatite found in previous studies of other disks was based on fitting of the short wavelength 5-17 $\mu$m Spitzer IRS spectra. This range contains both the strong 9.3 $\mu$m enstatite band, as well as weaker bands near 14-15 $\mu$m. These latter bands would be sensitive to the detection of enstatite in the spectrum. This can affect the outcome of our analysis. Thus, we also attempted to fit the MIRI spectrum from 5 $\mu$m to 13 $\mu$m, where a strong enstatite band appears at 9.3 $\mu$m regardless of the 14 $\mu$m bands. We find that enstatite remains undetected. In addition, we compared the 10 $\mu$m silicate band in the MIRI and Spitzer spectra with the Spitzer IRS observations of 9 Herbig disks in \cite{Juhasz_etal2010}. A linear continuum from 7.8 $\mu$m to 12.8 $\mu$m was subtracted from the 10 $\mu$m silicate band, which were then normalized to obtain the shapes, as shown in \fg{Herbig_silicate}. The black solid line represents the median of the 9 Herbig disks, with the shaded area indicating the standard deviation. Herbig disks are known to be enstatite-rich based on the prominent enstatite band at 9.3 $\mu$m \citep{Juhasz_etal2010}. In \fg{Herbig_silicate}, both MIRI and Spitzer spectra do not show the enstatite band as in the Herbig disks. Thus, we conclude that micron-sized dust grains in the inner PDS~70 disk are depleted in enstatite.
    \begin{figure}
    \centering
    \includegraphics[width=\linewidth]{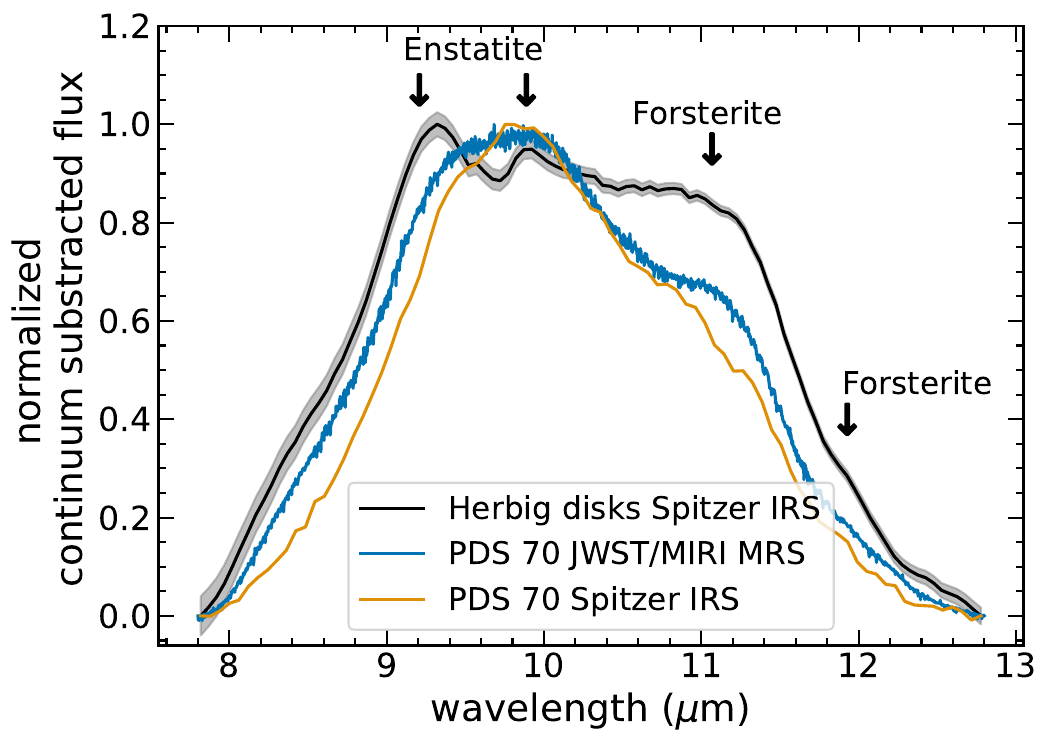}
    \caption{The 10 $\mu$m silicate band of the MIRI (blue) and Spitzer (orange) spectra of PDS~70 compared to 9 Herbig Ae/Be stars, observed by Spitzer IRS (gray). The arrows at 9.2 $\mu$m and 9.8 $\mu$m indicate the enstatite bands, while the arrows at 11 $\mu$m and 11.9 $\mu$m indicate the forsterite bands based on GRF opacities. }
    \label{fig:Herbig_silicate}
    \end{figure}

    \subsection{Variability of the inner disk dust}
    \label{sec:discuss_variability}
    \paragraph{}
    \cite{Gaidos_etal2024} show that the PDS~70 disk flux varies from $\sim$200 mJy to $\sim$1 mJy at 3.4 and 4.6 $\mu$m using NEOWISE photometry over $\sim$12 years, indicating that the innermost hottest dust can either appear or disappear entirely. Moreover, their photosphere-subtracted best-fit model suggest around 50~\% of the disk contribution around 5 $\mu$m, in agreement with \cite{Perotti_etal2023}. In both MIRI and Spitzer spectra, our fit shows the disk contribution to be $\sim 60$~\% of the flux, suggesting that we may be detecting PDS~70 during a phase where the inner disk contribution is transitioning between complete appearance and disappearance. The factor of 100 flux variation in this strong near-infrared variability is most easily explained by variations of the amount of optically thin hot dust in the innermost disk.

    As the bulk of hot dust appears and disappears, it provides a variable innermost disk opacity. This can affect the temperature of dust at larger distances and may provide a natural explanation of the temperature differences between the MIRI and Spitzer spectra. We can derive a rough distance of this warm dust using the following expression: 
    \begin{equation}
        \dfrac{T_{\rm dust}}{T_{\rm star}} = \left(\dfrac{r}{R_{\rm star}}\right)^{-0.5}
    \end{equation}
    and using the stellar temperature and the disk surface temperature derived above. This gives a distance of about 0.47 au. The Spitzer spectrum shows larger infrared excess around $5~\mu$m compared to the MIRI spectrum, which could indicate a more obscured innermost disk. Thus, the implication is that the temperature variations are probably not due to a difference in radial distance from the star, but are the result of varying irradiation of a stationary optically thin dust cloud. Due to this dust cloud in the innermost disk, radiation from the star is blocked, so the dust at larger radial distances can be cooler; the fitting results show that the disk surface temperature in the Spitzer spectrum is lower than in the MIRI spectrum in \fg{temperatures}.

    The range of midplane temperatures in both MIRI and Spitzer spectra are similar (third panel of \fg{temperatures}). The midplane temperature in Model Spitzer falls within the temperature range of Model GRF, and Model Mg+Fe is very similar to Model Spitzer at $\sim$140~K. Thus, the variability at wavelengths beyond 15 $\mu$m, where the optically thick midplane component in the model starts to contribute to the spectrum, is not due to a temperature difference, and does not follow the same trend as found for the optically thin dust component. One approach to increase the flux at longer wavelengths is to increase the amount of cold emitting dust.

     Despite the differences in temperatures and flux levels, the optically thin dust mass does not vary between MIRI and Spitzer spectra. In Model Spitzer, we found smaller dust grains compared to the MIRI spectra, and these smaller grains increase the dust surface area while maintaining the same dust mass. Dust evolution over the observation time from Spitzer to MIRI ($\sim$15 years) could possibly explain the variability, as the grain growth timescale is 1.8 years at 1 au ($\tau_{\rm grow} = 1/\epsilon\Omega_{\rm K}$ \citealt{Birnstiel_etal2016}, where $\epsilon$ is the dust-to-gas ratio of 0.1 in the inner disk \citealt{Portilla_etal2023}). Thus, the PDS~70 disk at the time of the Spitzer observation possibly had more cold and small dust, compared to the epoch of the MIRI observations.

    \subsection{Origin of the inner disk small grains}
    \label{sec:discuss_origin}
    \paragraph{}
    The PDS~70 disk likely represents the later stages of protoplanetary disk evolution, when accreting planets are formed. As the disk evolves, dust grains grow into larger bodies, such as pebbles or planetesimals, and dust gets depleted in the inner disk due to accretion onto the star. Due to the large gap in the PDS~70 disk and the resulting pressure bump in the outer disk ($\sim$77 au; \citealt{Pinilla_etal2024}), a continuous supply of dust from the outer disk is limited. Thus, the inner disk is expected to be dust-depleted, and the ALMA observation as well as detailed radiative transfer models indeed show a highly dust-depleted inner disk \citep{Benisty_etal2021,Portilla_etal2023}. However, mid-infrared observations of the PDS~70 disk show that the inner disk emission is strongly dominated by sub-micron-sized grains, more than most other disks observed so far. It is, therefore, interesting to discuss the possible origins of this abundant small grain component. We consider two scenarios to understand the origin of the small grains: 1) locally produced grains, either resulting from annealing or gas phase condensation or collisions among planetesimals, in the dynamical inner disk and 2) dust drifting inward from the outer disk. We discuss these two possibilities in the context of the dust composition derived in this study. 
    
        \subsubsection{Local production}
        \paragraph{}
        Small crystalline silicate grains can be locally produced in the inner disk as a result of the fragmentation of parent bodies or by gas phase condensation. In the case of gas phase condensation, the formation of both forsterite and enstatite can be expected if conditions for chemical equilibrium apply for a solar composition gas \citep{Woitke_etal2018}. This is not what we observe. There are several possibilities to suppress the formation of enstatite. Deviations from chemical equilibrium may result in the freeze-out of the reaction to form enstatite. Such an effect can, for instance, explain the formation of forsterite rather than enstatite during the flaring event in EX Lupi \citep{Abraham_etal2019}. A second way to avoid the formation of enstatite is by lowering the gas phase abundance of silicon \citep{Jorge_etal2021}.

        The small crystalline silicate grains can also be the result of parent-body processing and their release from these parent bodies through collisions. The presence of two gas giant planets in the outer gap and the highly variable inner-disk dust reservoir suggest that the PDS~70 disk is in an advanced stage of planet formation and inner-disk clearing. If the observed crystalline silicate grains in the inner disk are due to parent-body collisions, these parent bodies must be rich in forsterite. In addition, their low iron abundance suggests these parent bodies did not experience a high level of oxidation. Remarkably, \cite{Olofsson_etal2012} found a lack of enstatite from an analysis of small warm dust grains in debris disks. However, these grains contained significant amounts of iron. The grains we detect in the PDS70 disk, from a point of view of their iron content, are more in line with the iron-poor grains usually found in gas-rich protoplanetary disks. In the next subsection, we discuss another possibility for the origin of the crystalline silicates in the context of the pressure bump in the outer disk.

    \subsubsection{Filtration}
    \paragraph{}    
    The PDS~70 disk has a 54-au gap with two giant exoplanets, which produce a pressure bump at the outer edge of the gap. This pressure bump traps the drifting dust that is decoupled from the gas. However, small grains, which are more coupled to the gas, can penetrate the pressure bump and drift toward the inner disk. This is called filtration.
    
    Filtration at the outer edge of the gap may explain the presence of only forsterite for crystalline silicates in the disk. \cite{Pinilla_etal2024} model the dust evolution in the PDS~70 disk, using radiative transfer simulations to investigate the existence of an inner disk with the two giant planets in the gap. They suggest that small dust ($<0.1~\mu$m) is supplied by continuous fragmentation of trapped dust in the pressure bump. These small dust grains penetrate the pressure bump and supply dust to the inner disk. Although the inner disk of PDS~70 may have been initially enstatite-rich like other observed T~Tauri disks in \cite{Bouwman_etal2008, Olofsson_etal2010}, the enstatite could have been accreted into the central star due to a short accretion timescale. During this depletion of enstatite, small forsterite grains penetrate the pressure bump from the outer disk, which is forsterite-rich \citep{Bouwman_etal2008, Juhasz_etal2010, Olofsson_etal2010}, travel to the inner disk, and replenish the inner disk. Eventually, crystalline silicates in the inner disk could become dominated by forsterite. This process is schematically shown in \fg{schemFiltration}.

    \begin{figure*}
        \centering
        \includegraphics[width=\linewidth]{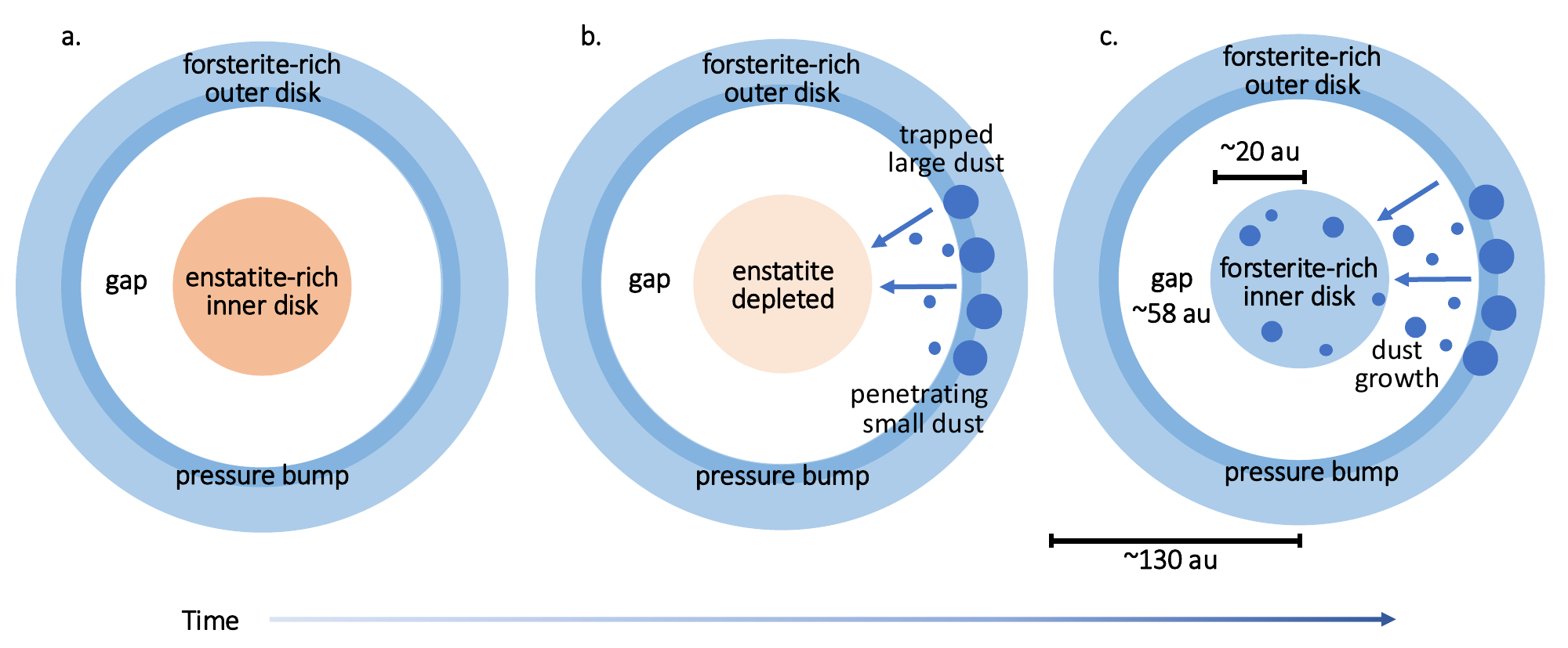}
        \caption{Sketch of the filtration mechanism in the PDS~70 disk. (a) The disk has a gap between the inner and outer disks. The inner disk is initially enstatite-rich, and the outer disk forsterite-rich. (b) While enstatite is depleted in the inner disk due to accretion onto the central star, small dust that penetrates the pressure bump travels through the gap to the inner disk. (c) The small dust supply the inner disk, and the inner disk eventually becomes forsterite-rich.}
        \label{fig:schemFiltration}
    \end{figure*}

    Observations indicate that outer disks are more forsterite-rich, but the origin of this mineralogical gradient is not well understood. \cite{Gail2004} in fact predicts an opposite trend, with an enstatite-rich outer disk resulting from the conversion of innermost forsterite to enstatite as small grains in the midplane are mixed outwards. The freeze-out of the reaction that converts forsterite to enstatite may prevent this process. Interestingly, the EX Lupi outburst reported by \cite{Abraham_etal2019} shows the production of only forsterite in the heated upper disk layers. These observations also indicate that the freshly formed forsterite was transported outward. If young stars experience frequent outbursts, this might provide a mechanism to enrich the outer disk with forsterite.

    While the penetrating small grains travel toward the inner disk, they grow to millimeter or centimeter-sized dust grains over 1 Myr \citep{Pinilla_etal2024}. The dust evolution models result in a power-law index of the grain size distribution as $\sim$3, indicating fewer small grains ($0.1~\mu$m) and more large grains ($\sim$10-100 $\mu$m) at 5, 10, 15 au, compared to the observed size distribution of interstellar grains (MRN distribution; \citealt{Mathis_etal1977}) at 10 au. These models suggest larger grain sizes than our best fit to MIRI spectrum. Theoretical simulations of grain growth through collisions \citep{Dominik_Dullemond2024} during its inward drifting \citep{Pinilla_etal2016,Pinilla_etal2024} suggest that grain sizes could grow to tens of microns or even centimeters.

    \cite{Dominik_Dullemond2024} show that the size distribution of dust grains becomes close to monodisperse around a few centimeters. On the other hand, \cite{Gaidos_etal2024} suggest a power-law index to be 3.5-4 based on 0.4m-LCOGT observations of the PDS~70 disk, which aligns with our finding of a few microns as the maximum grain size. The size distribution is smaller than what simulations predict, which only include collisional sticking, bouncing, and fragmentation. One possible explanation is fragmentation of icy grains after crossing the ice line. \cite{Pinilla_etal2024} do not consider the fragmentation of the grains into smaller sizes, leading to an overestimation of the grain size in the inner disk. 

    Throughout the dust growth process, grains may accumulate ices beyond their ice lines. When centimeter-sized icy pebbles cross the ice line, the ice sublimates, causing the dust aggregates to break apart into smaller dust grains. This process is described in the many-seeds scenario by \cite{Schoonenberg_Ormel2017}, and also suggested as a possibility for the PDS~70 disk in \cite{Perotti_etal2023}. \cite{Houge_etal2024} simulate grain growth in the many-seeds scenario and compare their results with ALMA observations of the V883 Ori disk. In their simulation, the ice line moves outward from 3 au to 80 au due to outburst events rather than dust drifting inward and passing the ice line. However, we can still discuss how icy dust responds when it crosses the ice line. In the many-seeds scenario, water ice trapped in dust sublimates, so the dust breaks apart into micron-sized grains. This change in grain size reduces the fragmentation barrier \citep{Birnstiel_etal2011}, which suppresses re-coagulation. Thus, the many-seed scenario may explain why the inner disk of the PDS~70 is dominated by sub-micron to micron-sized dust. 

    Warm H$_2$O has been detected in the MIRI spectrum and is well fitted by a slab model at 600 K within 0.05 au \citep{Perotti_etal2023}. \cite{Pinilla_etal2024} suggest that small grains can shield UV radiation, which prevents H$_2$O photodissociation \citep{Heays_etal2017}, and enriches the inner disk through the filtration mechanism. We note that \cite{Pinilla_etal2024} do not include UV photochemistry in their model. The water supply in the inner disk can be explained by the sublimation of ices in pebbles that have grown after penetrating the pressure bump and break apart at the ice line. A second possibility is the drift of oxygen gas through the gap, followed by efficient water formation in the warm and dense inner disk \citep{Perotti_etal2023, Portilla_etal2023}. \cite{Pinilla_etal2024} discuss that the initial inner disk is accreted within the first million years of the dust evolution, and the dust from the outer disk starts to replenish the inner disk in the 1-3.5 Myr range. Because the age of PDS 70 is estimated to be $\sim 5.4 \pm 1$ Myr \citep{Keppler_etal2018}, filtration and fragmentation may still be ongoing. This implies that the observed dust in the PDS 70 disk is more likely to be primitive rather than the result of parent-body processing.
    
    Finally, we note that both the filtration mechanism as well as collisions between parent bodies can contribute to the inner disk small grain population. Local production of crystalline silicates via flash heating or shock heating could enhance the crystallinity of small dust in the inner disk while depleting enstatite. While forsterite grains form locally in the inner disk, small forsterite grains from the outer disk penetrate the pressure bump and resupply the inner disk. These forsterite grains may fragment upon crossing the ice line and become sub-micron sized as we discussed above.

\section{Conclusion }
\label{sec:conclusion}
\paragraph{}
We have analysed the PDS~70 spectrum recently obtained by JWST/MIRI MRS, and archival Spitzer IRS spectrum, using the dust fitting model DuCK \citep{Kaeufer_etal2024}. DuCK employs a two-layer disk model described in \cite{Juhasz_etal2009,Juhasz_etal2010} and uses the MultiNest algorithm to identify the best model through retrieval with the dust mass fractions determined by the non-negative least squares fitting. We find that the MIRI spectrum can be fitted with a single temperature for the optically thin disk surface. We use three different types of absorption efficiencies: GRF, DHS, and Aerosol. The outcome of the fitting procedure is dependent on the choice of opacities, with GRF providing the best fitting model to the spectrum. Thus, this dependency should be kept in mind for future dust retrievals. We summarize our conclusions as follows:
\begin{itemize}
    \item Small amorphous silicates are the major dust species in the PDS~70 inner disk and dominate the spectra. The strength of the optically thin 10 $\mu$m silicate band is exceptionally strong, and any contribution of optically thick dust is minor at those wavelengths. This is different from most other observed disks.
    \item The model with GRF dust opacities provides the best fit to the observations. 
    \item The silicate dust in the PDS~70 disk is dominated by Mg-rich silicates. The analysis of the 23-24 $\mu$m band suggests around or less than $\sim$10 \% iron content in crystalline silicates, and dust fitting with both pure Mg silicates and Fe-rich silicates results in no detection of Fe-rich silicates or only in the large dust grains of Fe-rich silicate. The iron may be in metallic form, which our analysis is not sensitive to.
    \item In Model GRF and DHS, we find no evidence for enstatite, and Model Aerosol shows enstatite at 2 \%. Forsterite is detected in range of 4-20 \%. 
    \item The Spitzer spectrum, taken 15 years earlier, shows dust mineralogy similar to the MIRI spectrum, but the grain sizes are smaller and colder at roughly the same optically thin dust mass. This could indicate grain growth over the $\sim$15 years time span between the Spitzer and MIRI observations.
    \item The temperature difference of the optically thin silicate dust between MIRI and Spitzer spectra may be due to a change in the innermost disk opacity, resulting in variations of the flux received by the star at the same physical distance. This is likely due to a dynamic innermost disk, as also evidenced by the strong near-IR variability \citep{Gaidos_etal2024}. 
    \item Although the variability beyond 15 $\mu$m is not well understood and requires radiative transfer modeling, it could be due to higher abundances of small dust grains at large distances in Spitzer spectrum, which increases the emitting surface of cold dust. 
    \item The low abundance of enstatite and the iron-poor nature of crystalline silicates suggest that the small grains in the inner disk are more likely the result of the filtration and inward drift of small grains from the outer disk. Such grains may also transport water to the inner disk and fragment at the snow line. However, collisions of enstatite-depleted parent bodies and local heating processes can still produce the inner disk dust grains. 
\end{itemize}

Iron can exist in featureless dust species such as metallic iron or FeS, despite the iron-poor silicate dust species. Besides iron, other featureless dust species like amorphous carbon were also not accounted for this study. Investigating these species in the disk is a topic for future study, helping us to understand the iron and carbon content in planets. Moreover, the dust opacities depend on the provided optical constants. Expanding our study by including different sets of optical constants for GRF particle shapes could provide a better fit, especially near 8 $\mu$m. Because dust fitting results depend on the shape of opacities, different sets of optical constants should also be investigated.

\begin{acknowledgements}
This work is based on observations made with the NASA/ESA/CSA James Webb Space Telescope. The data were obtained from the Mikulski Archive for Space Telescopes at the Space Telescope Science Institute, which is operated by the Association of Universities for Research in Astronomy, Inc., under NASA contract NAS 5-03127 for JWST. These observations are associated with program \#1282. The following National and International Funding Agencies funded and supported the MIRI development: NASA; ESA; Belgian Science Policy Office (BELSPO); Centre Nationale d’Etudes Spatiales (CNES); Danish National Space Centre; Deutsches Zentrum fur Luft- und Raumfahrt (DLR); Enterprise Ireland; Ministerio De Econom\'ia y Competividad; Netherlands Research School for Astronomy (NOVA); Netherlands Organisation for Scientific Research (NWO); Science and Technology Facilities Council; Swiss Space Office; Swedish National Space Agency; and UK Space Agency. We acknowledge to Combined Atlas of Sources with Spitzer IRS Spectra (CASSIS) database for the use of low-resolution spectra. 

V. C. thanks the Belgian F.R.S.-FNRS, and the Belgian Federal Science Policy Office (BELSPO) for the provision of financial support in the framework of the PRODEX Programme of the European Space Agency (ESA) under contract number 4000142531.
G.P. gratefully acknowledges support from the Max Planck Society. 
E.v.D. acknowledges support from the ERC grant 101019751 MOLDISK and the Danish National Research Foundation through the Center of Excellence ``InterCat'' (DNRF150). 
T.H. and K.S. acknowledge support from the European Research Council under the Horizon 2020 Framework Program via the ERC Advanced Grant Origins 83 24 28. 
I.K., A.M.A., and E.v.D. acknowledge support from grant TOP-1 614.001.751 from the Dutch Research Council (NWO). 
B.T. is a Laureate of the Paris Region fellowship program, which is supported by the Ile-de-France Region and has received funding under the Horizon 2020 innovation framework program and Marie Sklodowska-Curie grant agreement No. 945298. 
D.G. thanks the Belgian Federal Science Policy Office (BELSPO) for the provision of financial support in the framework of the PRODEX Programme of the European Space Agency (ESA).
M.T. acknowledges support from the ERC grant 101019751 MOLDISK.
D.B. has been funded by Spanish MCIN/AEI/10.13039/501100011033 grants PID2019-107061GB-C61 and No. MDM-2017-0737. 
\end{acknowledgements}
    
\bibliographystyle{aa}
\bibliography{ref}

\begin{appendix}
\section{Additional dust retrieval results}
\paragraph{}
In this paper, we show multiple dust fitting models for MIRI spectrum with different sets of dust opacities, dust compositions, and also for Spitzer data. The mass fractions of each dust species and sizes in the models are summarized in Table \ref{tab:results}. The results of Model DHS and Model Aerosol are shown in \fg{DHSfitting} and \fg{Aerosolfitting}, respectively.

\begin{figure}[H]
\centering
\includegraphics[width=\linewidth]{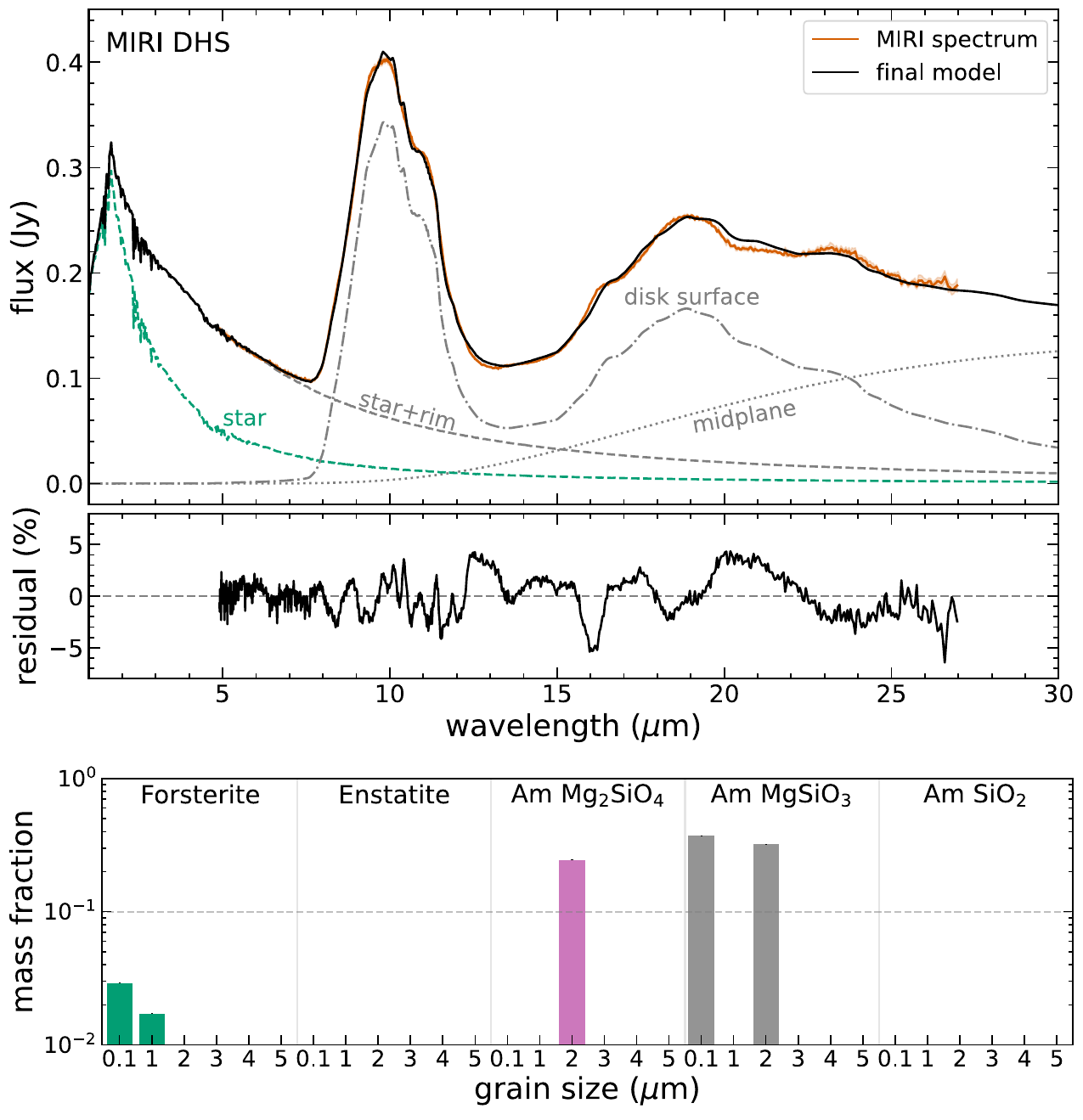}
\caption{Fitting with DHS opacity. The modeled spectrum generally matches well with the MIRI spectrum except the overestimation around 20 $\mu$m. Amorphous silicates still dominate the dust mass abundance, and enstatite and silica are not detected. }
\label{fig:DHSfitting}
\end{figure}

\begin{figure}
\centering
\includegraphics[width=\linewidth]{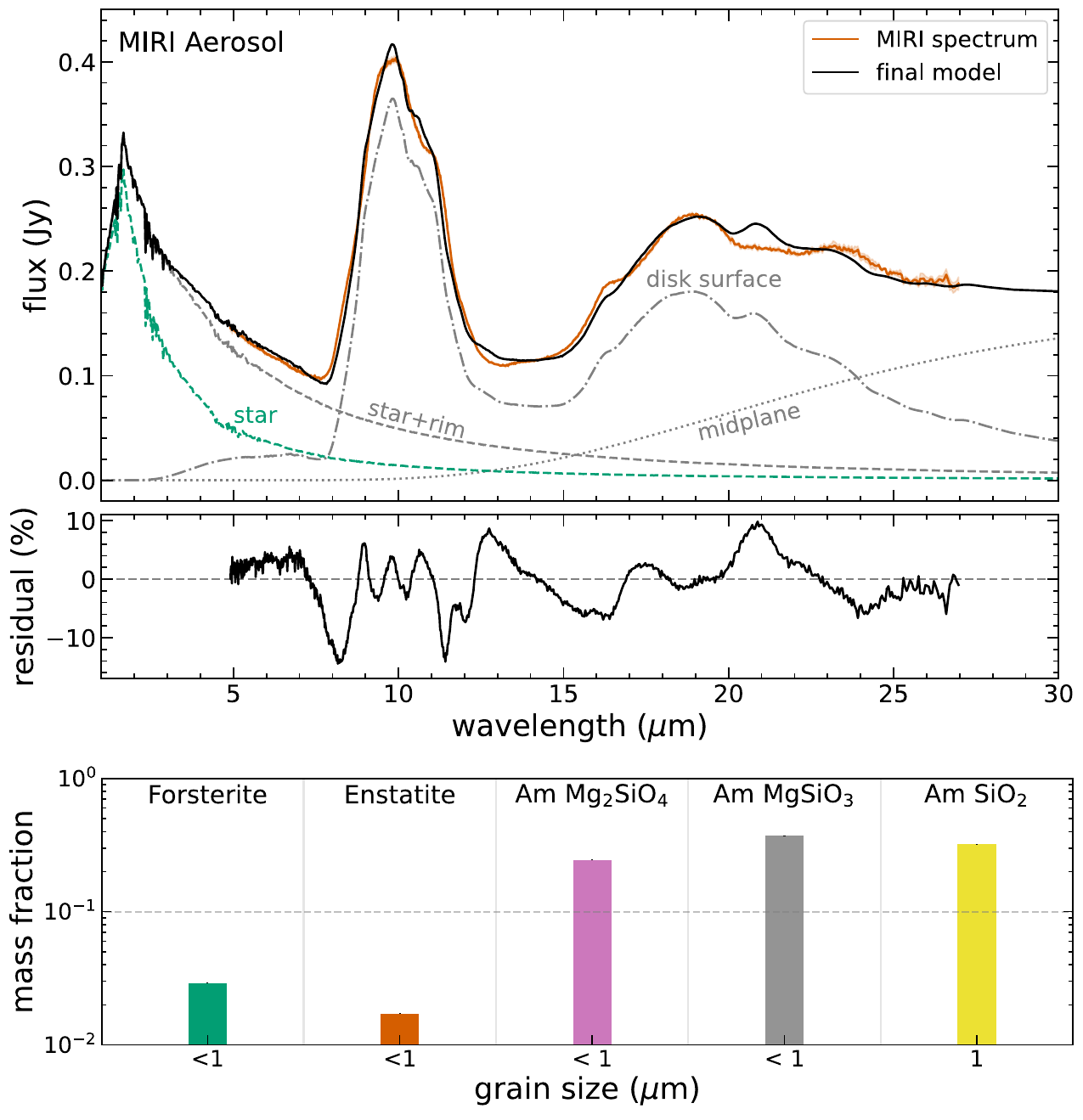}
\caption{Fitting with Aerosol measurements. The shape of the fitted model roughly matches with some mismatches at, for example, 17 $\mu$m and 22 $\mu$m, resulting large residuals. Amorphous silicates still dominate the mass fraction. Unlike from the models with GRF and DHS, enstatite and amorphous SiO$_2$ appear less than 10 \%.}
\label{fig:Aerosolfitting}
\end{figure}

\begin{figure}
\centering
\includegraphics[width=\linewidth]{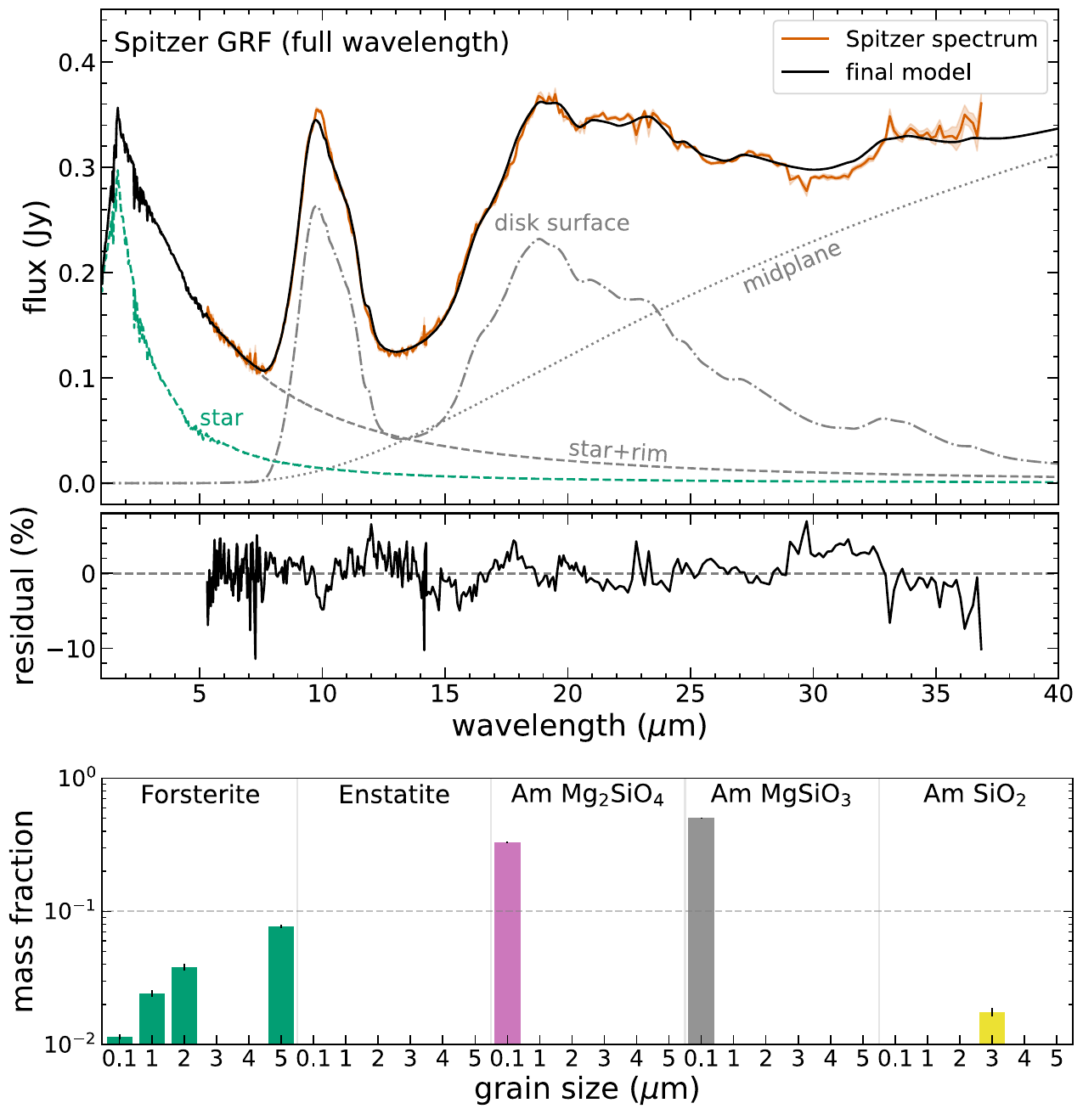}
\caption{Fitting of Spitzer data for full wavelength range (5-37 $\mu$m)}
\label{fig:SpitzerFull}
\end{figure}

\begin{table*}[]
\centering
\caption{Mass fraction of dust species and their grain sizes for each model. Most of errorbars is less than 0.01 \%, and the largest errorbar is still less than 0.2 \%. Thus, we do not denote these small errorbars in this table.}
\label{tab:results}
\begin{tabular}{cccccccccc}
\hline\hline
\multirow{2}{*}{name} & \multirow{2}{*}{grain model} & \multirow{2}{*}{species} & \multicolumn{7}{c}{mass fraction [\%]} \\
 &  &  & 0.1 $\mu$m & 1 $\mu$m & 2 $\mu$m & 3 $\mu$m & 4 $\mu$m &  5 $\mu$m & total \\ \hline
Model GRF & GRF & Forsterite & 0 & 5.7 & 2.7 & 0 & 0 & 12.0 & 20.4 \\
 &  & Enstatite & \textless{}1 & 0 & 0 & 0 & 0 & 0 & \textless{}1 \\
 &  & Amorphous Mg2SiO4 & 0 & 5.8 & 0 & 23.4 & 0 & 0 & 29.2\\
 &  & Amorphous MgSiO3 & 32.8 & 4.6 & 0 & 0 & 12.4 & 0 & 49.8\\
 &  & Amorphous SiO2 & 0 & 0 & 0 & 0 & 0 & 0  & 0\\
 \hline
Model GRF4$\mu$m & GRF & Forsterite & 0 & 5.8& 1.7 & 2.0 & 7.9 & - &17.4 \\
 &  & Enstatite & \textless{}1 & 0 & 0 & 0 & 0 & 0 & \textless{}1 \\
 &  & Amorphous Mg2SiO4 & 0 & 5.1 & 0 & 25.9 & 0 & 0 & 31.0 \\
 &  & Amorphous MgSiO3 & 33.9 & 4.5 & 0 & 0 & 1.3 & 0 & 39.7\\
 &  & Amorphous SiO2 & 0 & 0 & 0 & 0 & 0 & 0 & 0\\
 \hline
Model GRF2$\mu$m & GRF & Forsterite & 0 & 5.0 & 7.2 & - & - & - & 12.2\\
 &  & Enstatite & \textless{}1 & 0 & 0 & 0 & 0 & 0 & \textless{}1\\
 &  & Amorphous Mg2SiO4 & 0 & 0 & 19.3 & 14.7 & 0 & 0 & 34.0\\
 &  & Amorphous MgSiO3 & 39.2 & \textless{}1 & 0 & 0 & 13.7 & 0 & 52.9\\
 &  & Amorphous SiO2 & \textless{}1 & 0 & 0 & 0 & 0 & 0 & \textless{}1\\
 \hline
Model DHS & DHS & Forsterite & 2.9 & 1.7 & 0 & 0 & 0 & 0 & 4.6\\
 &  & Enstatite & 0 & 0 & 0 & 0 & 0 & \textless{}1 & \textless{}1\\
 &  & Amorphous Mg2SiO4 & 0 & 0 & 24.5 & 0 & 0 & \textless{}1 &24.5\\
 &  & Amorphous MgSiO3 & 37.2 & 0 & 32.3 & 0 & 0 & 0 & 69.5\\
 &  & Amorphous SiO2 & 0 & \textless{}1 & 0 & 0 & 0 & 0 & \textless{}1\\
 \hline
Model Aerosol &  & Forsterite & - & 3.2 & - & - & - & - & 3.2\\
 &  & Enstatite & - & 2.1 & - & - & - & - & 2.1\\
 &  & Amorphous Mg2SiO4 & - & 61.2 & - & - & - & - &61.2 \\
 &  & Amorphous MgSiO3 & - & 26.0 & - & - & - & - & 26.0\\
 &  & Amorphous SiO2 & - & 7.4 & - & - & - & - & 7.4\\
 \hline
Model Mg+Fe & GRF & Forsterite &  0  & 4.7 &4.7  &0  &0  &4.5 & 13.9\\
 &  & Enstatite &  <1 & 0 & 0 & 0 & 0 & 0 & \textless{}1\\
 &  & Fayalite &  <1 &  & 0 & 0 & 0 & 0 &\textless{}1\\
 &  & Amorphous Mg2SiO4 & 12.9 & 0 & 0 & 0 & 0 & 0 &12.9\\
  &  & Amorphous MgFeSiO4 & 0 & 0 & 0 & 0 & 11.4 & 9.2 &20.6\\
 &  & Amorphous MgSiO3 &  27.0&8.11  & 0 & 0 & 0 & 0 &35.11\\
 &  & Amorphous MgFeSi2O6 & 0 & 0 & 0 & 0 & 0 & 15.9  & 15.9 \\
 &  & Amorphous SiO2 & 0 & 0 & 0 & 0 & 0 & 0 & 0\\
 \hline
Model Spitzer & GRF & Forsterite & 0 & 3.7 & 6.1 & 0 & 0 & 0 & 9.8\\
 &  & Enstatite & 0 & 0 & 0 & 0 & 0 & 0 & 0\\
 &  & Amorphous Mg2SiO4 & 36.0 & 0 & 0 & 0 & 0 & 0 & 36.0\\
 &  & Amorphous MgSiO3 & 37.6 & 12.7 & 0 & 0 & 0 & 0 &50.3\\
 &  & Amorphous SiO2 & 0 & 0 & 0 & 0 & 0 & 3.8 & 3.8\\
 \hline
Model Spitzer(full) & GRF & Forsterite & \textless{}1 & 3.2 & 2.7 & 0 & 0 & 8.6 &14.5\\
 &  & Enstatite & 0 & 0 & 0 & 0 & 0 & 0 & 0 \\
 &  & Amorphous Mg2SiO4 & 33.4 & 0 & 0 & 0 & 0 & 0 &33.4\\
 &  & Amorphous MgSiO3 & 49.7 & 0 & 0 & 0 & 0 & 0 & 49.7\\
 &  & Amorphous SiO2 & 0 & 0 & 0 & 1.6 & 0 & 0 & 1.6\\ \hline
\end{tabular}
\end{table*}

\section{Posterior probability distribution}
\paragraph{}
\Fg{GRFposterior} is the posterior probability distributions of Model GRF, the best model for MIRI data. The temperatures of each component are derived through Bayesian analysis, while scale factors and dust abundances are estimated using the non-negative square fitting. Circular contours suggest that results are independent from other parameters. In contrast, tilted elliptical contours indicate certain dependency between components. For example, the 0.1 $\mu$m-sized amorphous MgSiO$_3$ and surface temperature exhibit a degeneracy. As the temperature decreases, the abundance increases, which is natural to balance the flux contribution. Although some other degenerate relations do not follow this trend, they vary within very small range, demonstrating that abundances and temperatures are well-constrained. 

\begin{figure*}
\centering
\includegraphics[width=\linewidth]{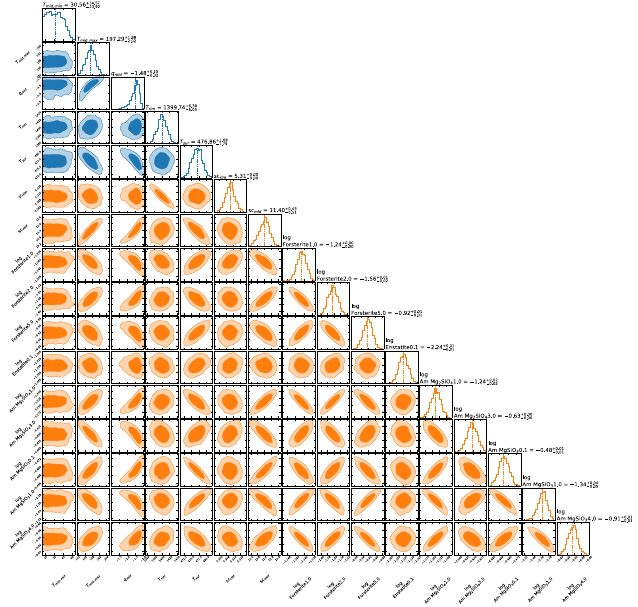}
\caption{Posterior probability distributions of Model GRF. The scale factors for inner rim (sc$_{\rm rim}$) and midplane (sc$_{\rm mid}$) components and dust abundances are estimated by the non-negative least square fitting (orange), while the other parameters are determined by Bayesian analysis (blue). The error bars represent the 16th and 84th percentile that correspond to the -1 sigma and 1 sigma uncertainties of the retrieved values. }
\label{fig:GRFposterior}
\end{figure*}

\section{Shifting the 23-24 $\mu$m band peak}
\paragraph{}
The peak position of 23-24 $\mu$m band does not only depend on the relative fraction of iron with respect to magnesium but also on grain size. As the grain is larger, the peak position shifts to the longer wavelength. For 0.1 $\mu$m, 2 $\mu$m, and 3 $\mu$m, the peak positions shift from 23.27 $\mu$m to 23.46 $\mu$m and 23.66 $\mu$m as shown in \fg{23micron_shift}. 
\begin{figure}
    \centering
    \includegraphics[width=\linewidth]{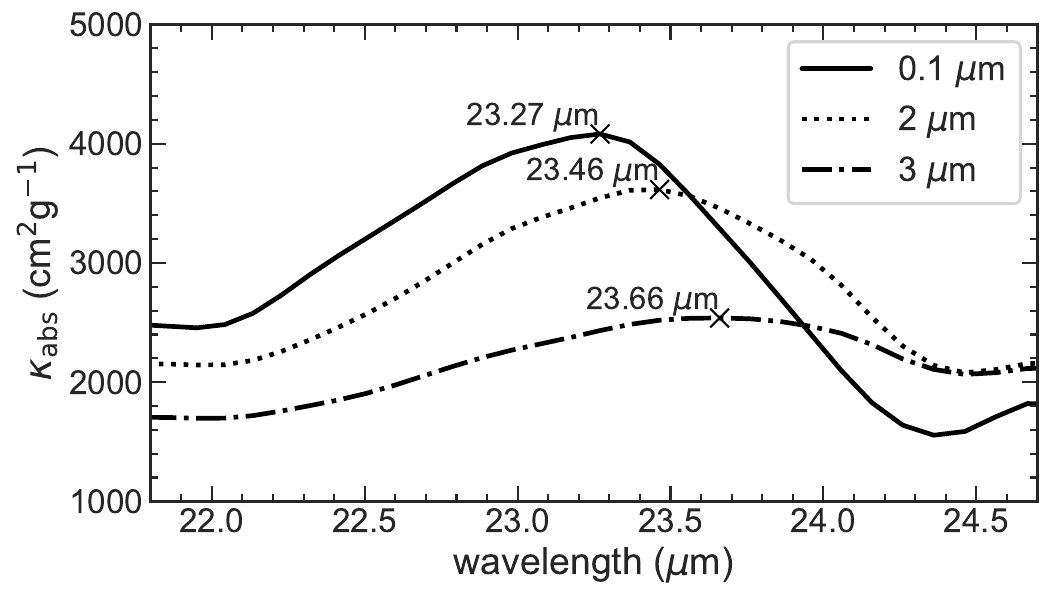}
    \caption{23-24 $\mu$m crystalline features of different grain sizes in GRF opacities. As the grain sizes become large, the peak shifts to longer wavelength.}
    \label{fig:23micron_shift}
    \end{figure}

\end{appendix}

\end{document}